\documentclass{JINST}

\title{Performances of a GEM-based Time Projection Chamber prototype for 
the AMADEUS experiment}

\author{M.~Poli Lener$^a$\thanks{Corresponding author.}~, M.~Bazzi$^a$, G.~Corradi$^a$, C.~Curceanu$^a$, A.~D'Uffizi$^a$,
 C.~Paglia$^a$, A.~Romero Vidal$^a$~$^b$, E.~Sbardella$^a$, A.~Scordo$^a$, 
D.~Tagnani$^a$, J.~Zmeskal$^c$\\
\llap{$^a$}Laboratori Nazionali di Frascati -INFN,\\
  Frascati, Italy\\
\llap{$^b$}Universidad de Santiago de Compostela,\\
  Santiago de Compostela, Spain\\
\llap{$^c$}Stefan Meyer Institut fur s\"{u}batomare Physik,\\
  Vienna, Austria\\
  E-mail: \email{marco.polilener@lnf.infn.it}}

\abstract{A large number of high-energy and heavy-ion experiments successfully 
used Time Projection Chamber (TPC) as central tracker and particle 
identification detector. However, the performance requirements on TPC for new
high-rate particle experiments greatly exceed the abilities of traditional
TPC read out by multi-wire proportional chamber (MWPC). Gas Electron
Multiplier (GEM) detector has great potential to improve TPC performances
when used as amplification device.\\
In this paper we present the R\&D activity on a new GEM-based TPC  
detector built as a prototype for the inner part for AMADEUS, a new 
experimental proposal at the DA$\Phi$NE collider at Laboratori Nazionali di 
Frascati (INFN), aiming to perform measurements 
of the low-energy negative kaons interactions in nuclei.\\
In order to evaluate the GEM-TPC performances, a 10x10 cm$^2$ prototype with a 
drift gap up to 15 cm has been realized. The  
detector was tested at the $\pi$M1 beam facility of the Paul Scherrer 
Institut (PSI) with low momentum pions and protons, without magnetic field. 
Drift properties of argon-isobutane gas mixtures are measured and 
compared with Magboltz prediction.
Detection efficiency and spatial resolution as a function of a large 
number of parameters, such as the gas gain, the drift 
field, the front-end electronic threshold and particle momentum, are 
illustrated  and discussed. Particle identification capability and the 
measurement of the energy resolution in isobutane-based gas mixture are 
also reported.}


\begin{document}

\section{Introduction}
\label{sec:intro}
\begin{figure}[!ht]
 \centering
    \includegraphics[width=7.9cm]{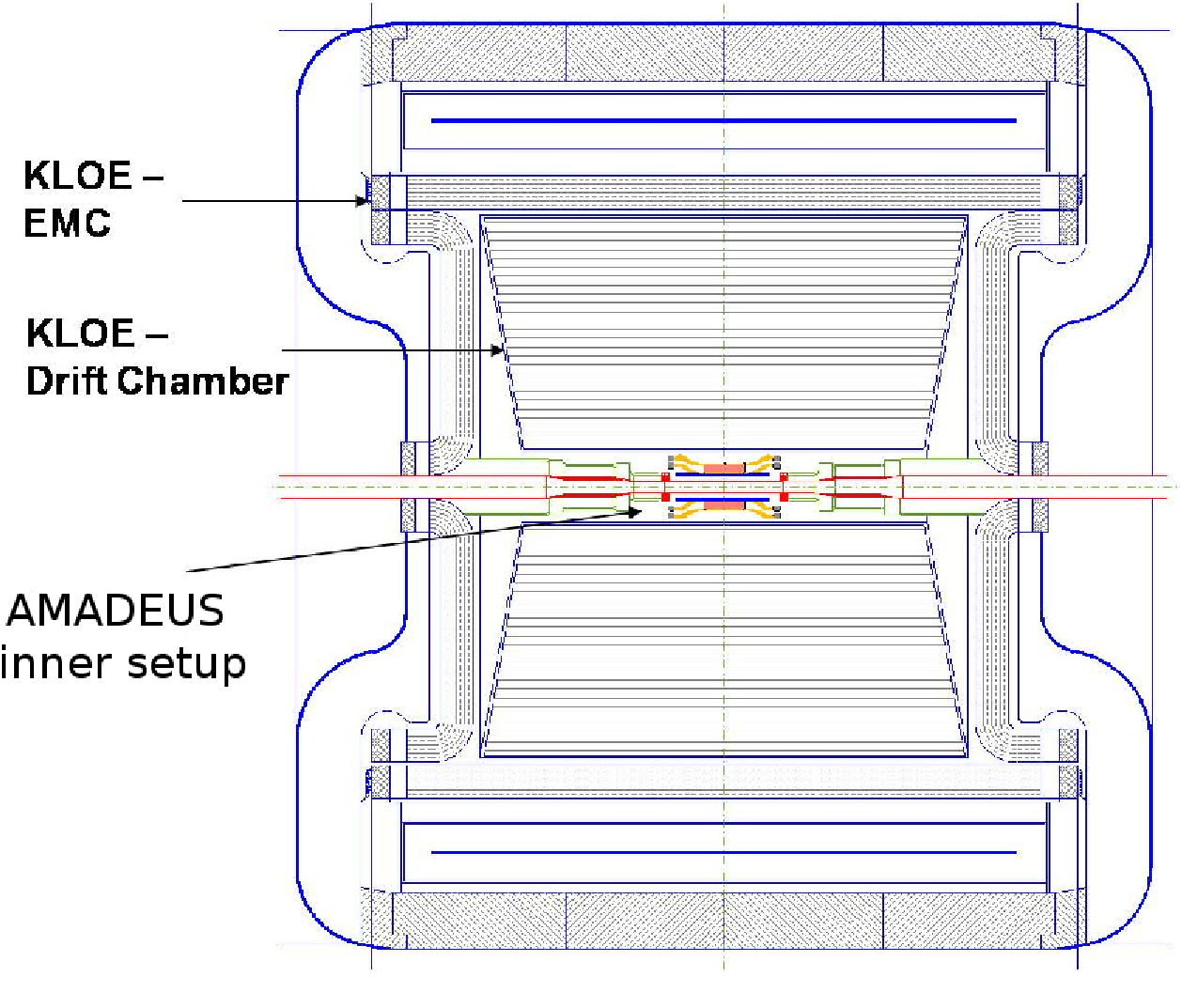}
    \caption{Cross-section of the KLOE detector including the 
      AMADEUS inner setup inside the Drift Chamber.}
    \label{fig:amadeus}
\end{figure}

\begin{figure}[!ht]
  \centering
    \includegraphics[width=7.5cm]{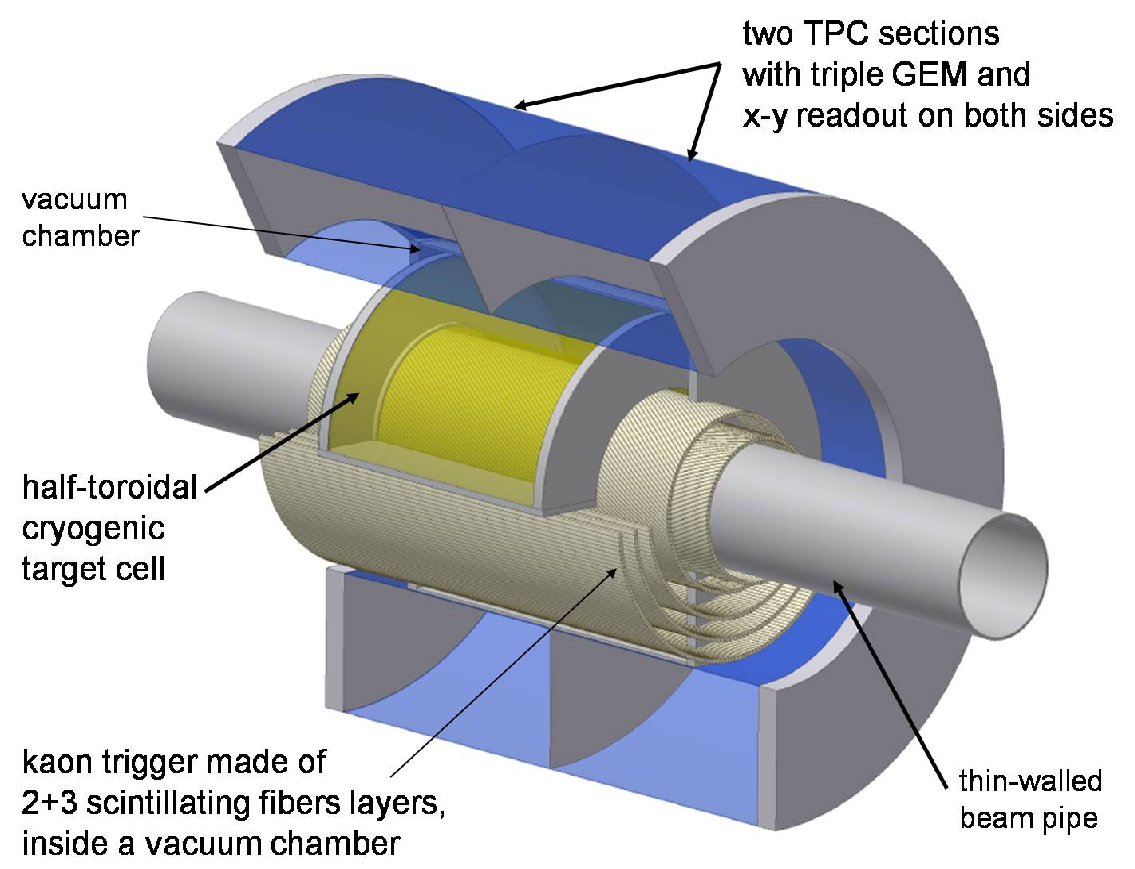}
    \caption{The inner setup of AMADEUS. From the beam pipe to
        the outer region: the Kaonic Trigger, the Gas Target and the TPG.}
    \label{fig:target}
\end{figure}

The AMADEUS experiment~\cite{AMADEUS},\cite{AMADEUS1} aims to perform 
measurements of low-energy 
charged kaons interaction in nuclear matter, in particular to search for the 
so-called ``kaonic nuclear clusters''. The AMADEUS setup is going to be 
installed inside the KLOE detector~\cite{KLOE} in the free space inside the 
drift chamber (Fig.~\ref{fig:amadeus}). The experiment will then use the 
drift chamber (DC) and the calorimeter of the KLOE detector, together with a 
dedicated setup consisting of a target cell to be filled with deuterium or 
helium (3 and 4), a trigger system, which will trigger on the back-to-back 
K$^+$ K$^-$ pairs emitted from the decay of the $\Phi$
particles produced at the  DA$\Phi$NE  e$^+$ e$^-$ collider of the 
LNF-INFN~\cite{KLOE1},\cite{KLOE2},\cite{scordo} and an inner tracker, 
namely a Time Projection Chamber, TPC, equipped with Gas Electron 
Multiplier~\cite{sauli} (Fig.~\ref{fig:target}).\\    
%The change of the hadron masses and hadron interaction in the nuclear
%medium and the structure of cold dense hadronic matter are hot topic of
%hadron physics today. These important problems, yet unsolved, will be the
%research field of the AMADEUS experiment~\cite{AMADEUS}. AMADEUS will perform
%the first complete experimental study of the so-called $Kaonic$
%$Clusters$. Such a study has deep consequences in a still open sector of the
%strangeness hadronic/nuclear physics: how the hadron masses and hadron
%interactions change in the nuclear medium with consequences on the
%structure of cold dense hadronic matter.\\
%AMADEUS will search for
%antikaon-mediated deeply bound nuclear states, which could represent,
%indeed, the ideal condition for investigating the way in which the
%spontaneous and explicit chirical symmetry breaking pattern of low-energy
%QCD occur in the nuclear environment. The existence or not of deeply bound
%systems is presently matter of vivid discussion among theoreticians and
%experimentalists.
%The AMADEUS setup will be within the Drift Chamber (DC) of KLOE 
%(Fig.~\ref{fig:amadeus}), where a free area with a diameter of 50 cm 
%is available, and it will consist
%of a Gaseous Target, a Kaonic Trigger (Scintillator Fiber and 
%Silicon Photon Multiplier readout) 
%and a Time Projection Chamber equipped with Gas
%Electron Multiplier~\cite{sauli} as Inner Tracker (Fig.~\ref{fig:target}).\\
The GEM-based TPC (TPG) will be 20 cm long with an inner diameter of 8 cm and
 an outer one of 40 cm.\\
The required performances for the TPG AMADEUS are: a spatial
resolution better than 200 $\mu$m in X-Y and 300 $\mu$m in Z in order to
achieve together with the KLOE DC a momentum resolution better than 1$\%$ in a magnetic
field of 0.5 T; a detector material budget lower than 0.5$\%$ of X$_0$; a rate
capability of $\sim$ 1 kHz/cm$^2$~\cite{MPL}. In addition, the detector must
operate in continuous mode, which means an ion feedback below
10$^{-3}$~\cite{sauli1}, and it must tolerate, without performance losses, an
integrated charge of $\sim$~0.05 $C/cm^2$ in 1 year of operation at a gas
gain of 6000 and an average particle flux of 1~kHz/cm$^{2}$~\cite{desimone}.\\
Since most of the above requirements are easily fulfilled by a TPG, 
the R\&D activity at the Laboratori Nazionali di Frascati (INFN) is 
mainly focused on
the choice of the gas mixture, in order to achieve the highest spatial 
resolution with a 0.5 T value of magnetic field, and on the design of the
detector readout.\\
A prototype of the AMADEUS TPG detector was built and tested both in laboratory 
and at Paul Scherrer Institut (PSI). 
The technique used in the prototype construction is described in 
Sec.~\ref{sec:detector}, while the experimental setup 
is briefly reported in 
Sec.~\ref{sec:setup}. The choise of the gas mixtures is described in
Sec.~\ref{sec:gas} together with the estimate of the primary ionization 
and the measurement of the gas gain. In Sec.~\ref{sec:results} 
the overall detector performances obtained at the PSI, in terms of 
detection efficiency, spatial resolution, PID capability and energy resolution,
are presented and discussed. The paper ends with the conclusions.
%
%
%\begin{figure}[hb]
%  \centering
%    \includegraphics[height=3.5cm, width=4.cm]{amadeus.eps}%
%    %\qquad
%    \qquad
%    \includegraphics[height=3.5cm, width=4.cm]{target.eps}
%    \caption{Measurements of the electron drift velocity of the
%    Ar/CO$_2$/CF$_4$ (45/15/40) at STP:
%     up) moving the detector along z; down) changing the drift field.}
%    \label{fig:scanz}
%\end{figure}
%
%
\section{AMADEUS TPG prototype production}
\label{sec:detector}
\begin{figure}[!ht]
  \centering
    \includegraphics[width=7.5cm]{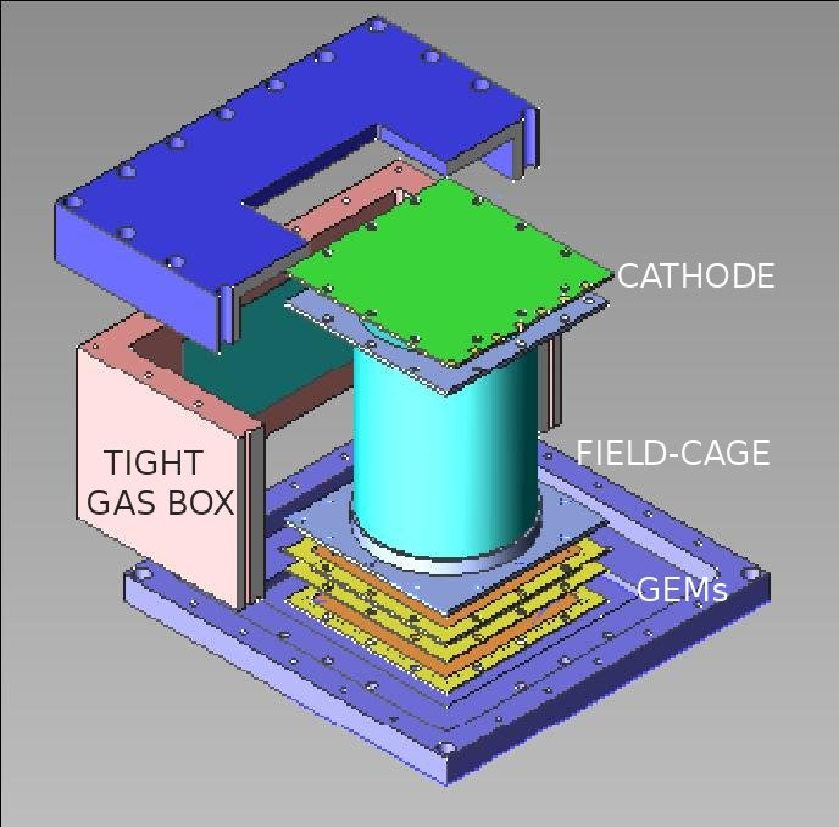}
    \caption{Exploded view of the GEM-based TPC detector.}
    \label{fig:detector1}
\end{figure}

The TPG prototype construction has been performed in a class 1000
clean room. The detector is composed by three GEM foils glued
on Fiberglass (FR4) frames, defining the gaps between GEM
themselves, then sandwiched between a cathode and anode
PCBs. The exploded view of the detector is shown in Fig.~\ref{fig:detector1}.

\begin{figure}[ht!]
  \centering
    \includegraphics[width=7.5cm]{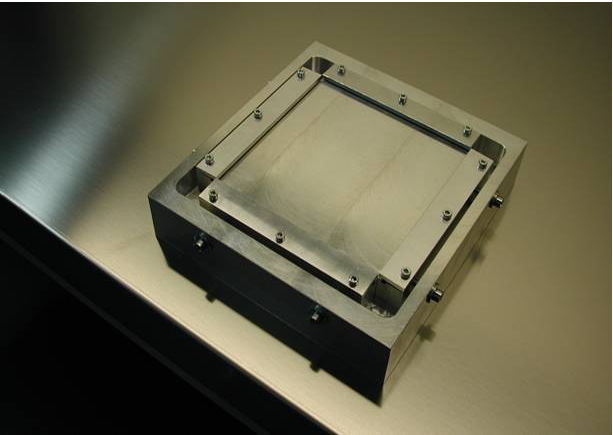}
% \caption{fig:stretcher}
    \caption{Picture of the home-made tool used for stretching the GEMs
      before the frame gluing. A mechanical tension of 1 kg/cm$^2$ is applied
      on the edges of the foil.}
    \label{fig:stretcher}
\end{figure}

The GEM foils, manufactured by CERN-EST-DEM workshop,
have an active area of 10x10 cm$^{2}$. Severe criteria have been adopted for GEM
foils acceptance: the defects, detected by means of microscope
inspection, must have an area smaller than 1 mm$^{2}$; 
%(Fig.~\ref{fig:difetti});
a maximum leakage current of 1 nA at 600 V, 
measured with a residual humidity of about 20\%, is required.\\
To ensure high stability of detector operation without
requiring a grid spacer inside the active area, a stretching 
technique of the GEM foil before the frame gluing has been 
used~\cite{stretcher}.
This technique was developed for the R\&D of the GEM for the LHCb Muon Chamber.
In Fig.~\ref{fig:stretcher} the home-made tool used for the production 
is shown.\\  
In Ref.~\cite{MPL} the extensive campaign of measurements is reported, 
such as the maximum sag due to electrostatic forces between GEMs, 
the kapton creep and radiation effects, which   
have been performed in order to avoid electrostatic instability 
and to achieve a good uniformity response.

\begin{figure}[h!]
  \centering
    \includegraphics[width=7.5cm]{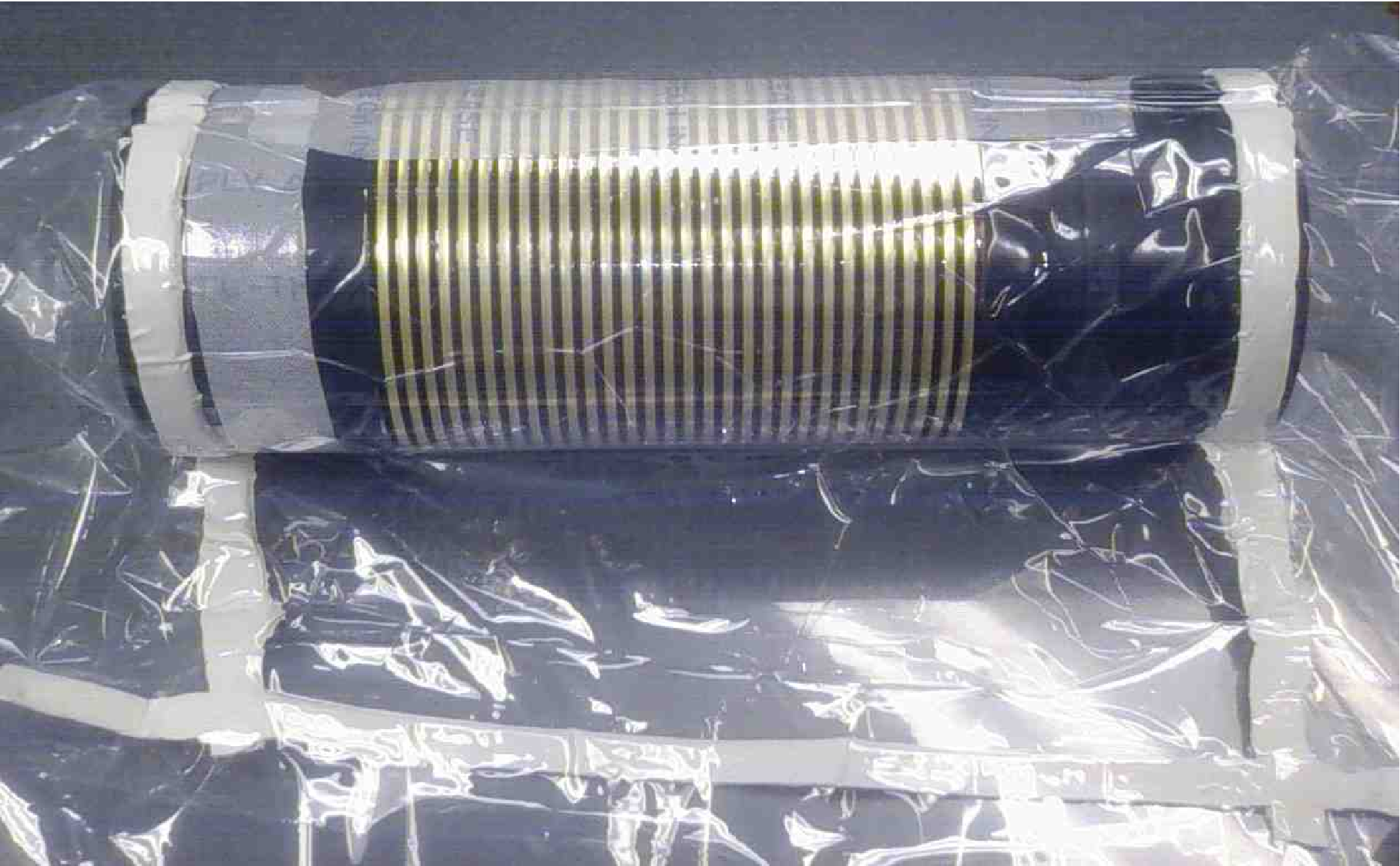}
    \caption{A cylindrical field-cage foil realised with vacuum bag technique.}
    \label{fig:fieldcage}
\end{figure}

\begin{figure}[ht!]
  \centering
    \includegraphics[width=7.5cm]{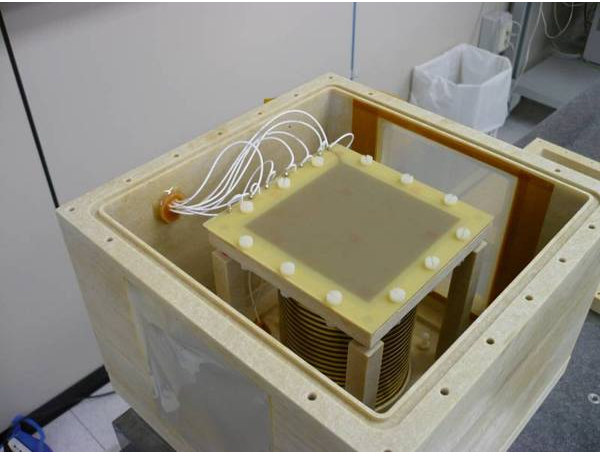}
    \caption{Tight gas box for the TPG. The two windows used for
             the particle crossing the detector and the HV connections of the
             various electrodes are visible.}
    \label{fig:detector}
\end{figure}

The electric field uniformity in the drift volume is provided by a
cylindrical field cage, which consists of two sets of copper strips 
(2.5 mm wide) on both sides of an insulating Kapton foil (15 cm height),
 where the outer strips cover the gaps between the inner strips.
The potential on each ring is defined by a precision
resistor chain located outside the gas volume.
The cylindrical shape has been obtained by exploiting
the vacuum bag technique~\cite{cilindrica} and rolling the copper-kapton foil
onto machined Polytetrafluorethylene (PTFE) cylinders that act as molds 
(Fig.~\ref{fig:fieldcage}).
The edges of the foil are glued together with bi-component epoxy 
along the axial millimetric overlap (2${-}$3 mm wide).
After the curing cycle, the field-cage foil is easily removed from 
the cylindrical mold: the results is a perfectly cylindrical field-cage.
To ensure a mechanical support, two fiberglass flanges are glued on the 
edges with Araldite 2011.\\ 
In this R\&D phase, the detector has been encapsulated inside a tight gas
box which has been provided by two 14x12 cm$^{2}$ windows, covered by 
10 $\mu$m  Mylar film, in order to reduce as much as possible the radiation 
length for a particle crossing the detector (Fig.~\ref{fig:detector}). 
%This solution is allows to simply change the detector geometry or to replace new GEM foil.
The gas tightness of the detector has been measured and it is lower than
 2 mbar per day, corresponding to less than 100 ppmV of residual humidity with 
a gas flux of 100 cc/min.

\section{Experimental test setup and data taking}
\label{sec:setup}
\begin{figure}[!ht]
  \centering
    \includegraphics[width=6.cm]{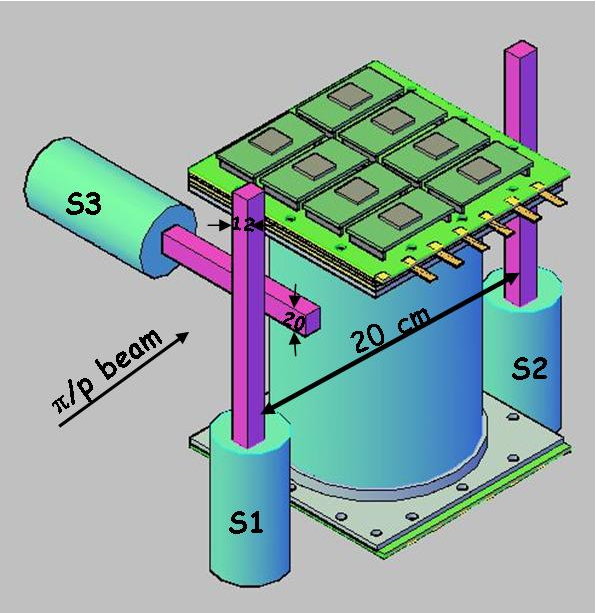}
    \caption{Schematic drawing of the experimental setup.}
    \label{fig:setup}
\end{figure}

The performance of the TPG prototype
has been studied at the $\pi$M1 beam facility of the Paul Scherrer Institut
(PSI) without magnetic field. The $\pi$M1 beam is a quasi-continuous 
high-intensity secondary beam providing pions or protons with very precise 
momentum resolution.\\
%up to $\sim $ 10$^7$ $\pi^-$/s ($\sim $ 10$^8$ $\pi^+$/s)  
%at 350 MeV/c or $\sim $ 10$^7$ proton/s at 500 Mev/c for each mA 
%of beam current in the primary beam. 
The study of the detection efficiency, spatial and energy resolution of the
detector has been performed with a beam rate
of $\sim$ 200 Hz.\\
The trigger consisted of the coincidence of three 
scintillators S1$\otimes$S3$\otimes$S2, centered on the beam axis.
The distance between the two scintillators S1$\otimes$S3 
and the S2 scintillator was about 20 cm as shown in Fig.~\ref{fig:setup}. 
The area covered by the intersection of these 3 scintillators is 
approximately 12x20 mm$^2$.\\ 
The horizontal size (12 mm) of S1 and S2 was chosen in order to completely
 cover the width of the detector readout as described in the following section, 
while the 20 mm width of the S3 scintillator was used in order to limit 
the vertical size of the beam. Such effect will be
illustrated in Sec~\ref{sec:drift}.\\
The coincidence of the discriminated S1, S2, S3 signals
was delayed to give the common stop to a 17-bit multihit CAEN TDC, 
with 0.1 ns resolution and 5 ns double edge resolution.\\
Moreover, another scintillator, about 5 m far from the S1, S2, S3 cross, was 
acquired in order to perform a measurement of the particle momentum crossing 
the setup by means of time of flight. The measured momentum resolution for 
pion and proton beams
in the momentum range between 100 and 440 MeV/c was less then 1\%.\\ 
In the following, unless otherwise stated, the beam was directed
perpendicular to the drift axis of the TPG prototype.
The discriminator threshold on the front-end electronics 
of the TPG signal was set to $\sim$ 3.5 fC and $\sim$ 5 fC.

\subsection{The readout electronics}
The prototype readout is composed by 4 rows of 32 pads for a total of 128 pads.
Each pad of $\sim$ 3x3 mm$^2$ was connected to a front-end
 board based on CARIOCA-GEM chip~\cite{carioca}. Such chip was developed for
the GEM readout in the LHCb Muon Apparatus and it allows to readout half of 
a pad row. 
The discriminated and digitized signals were sent to the 17-bit multihit
CAEN TDC. For each electronics channel the leading edge
(time hit) and the trailing edge, which allow to measure the Time Over 
Threshold (TOT) with respect to leading edge, were recorded. 
This TOT technique has been investigated in order to study the 
possibility to perform a charge measurement needed for 
particle identification, taking into account the advantage of a faster
data acquisition, pattern recognition and a better compression 
into track segment.   
%
%\subsection{Data Analysis}
%
\section{Gas Mixture Choice}
\label{sec:gas}
\begin{figure}[!ht]
  \centering
    \includegraphics[width=7.5cm]{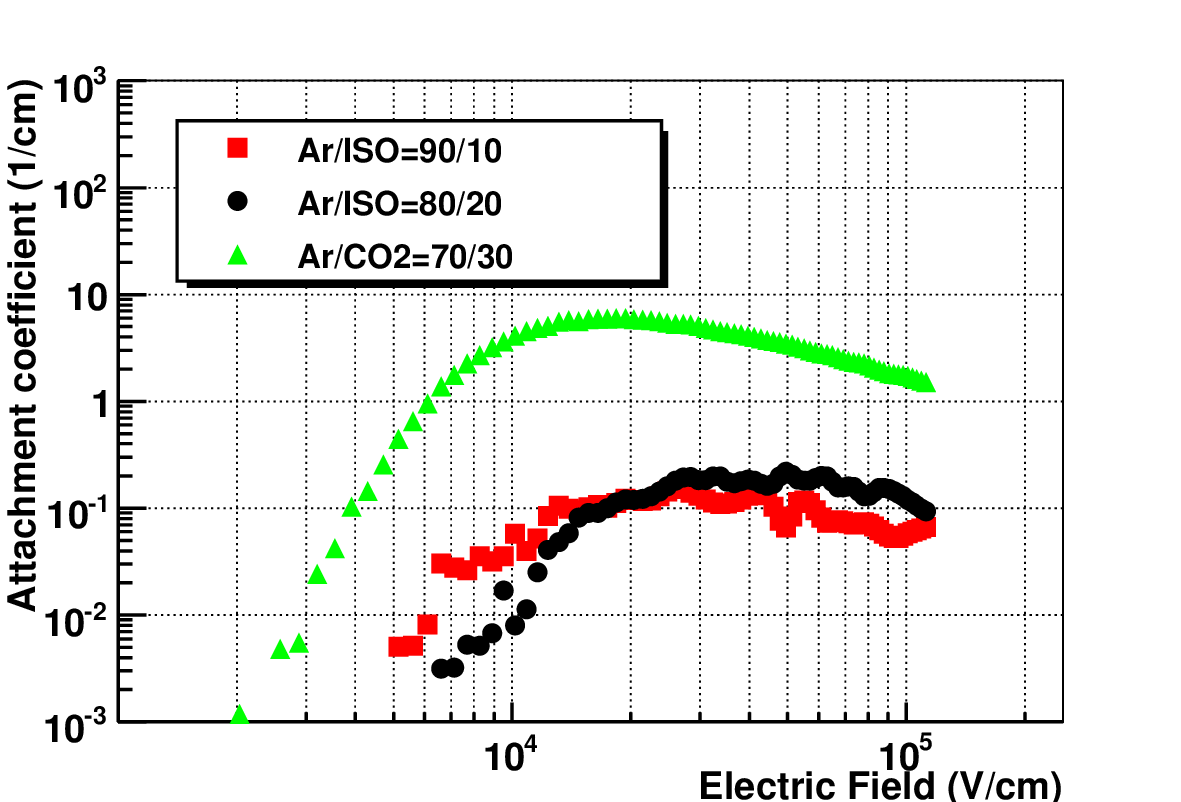}
    \caption{Attachment coefficient as a function of
             the electric field. The curve of the Ar/CO$_2$=70/30 
             gas mixture is reported for comparison.}
    \label{fig:att}
\end{figure}

During this R\&D phase we decided to use isobutane-based gas mixtures
since they are characterized by a large primary ionization 
(see Sec.~\ref{sec:cluster}), a high 
drift velocity (see Sec.~\ref{sec:drift}), a high Townsend coefficient, 
which allow to work at lower HV values (see 
Sec.~\ref{sec:gain}), a moderate longitudinal and transversal 
diffusions ($<$ 400 $\mu$m/$\sqrt{cm}$ for a drift field of 150 V)  
and last but not least a very low attachment coefficient as shown in 
Fig.~\ref{fig:att}.

\subsection{Primary Ionization}
\label{sec:cluster}

The primary ionization of the gas mixtures used during the beam test
has been estimated with GARFIELD~\cite{garfield} simulation tool, which
is the common framework used for the simulation of gas detectors.\\
In Tab.~\ref{tab:cluster} have been reported the number of clusters per cm
and the number of electrons per cluster for different momentum beam and for
the tested gas mixtures. It is worth noticing that the use of isobutane-based
gas mixtures allow to reach a larger primary ionization with respect to
the Ar/CO$_2$=70/30 gas mixture.

\begin{table}[!ht] 
\centering \scriptsize 
%\tiny   
\begin{tabular}{c c |c c c} 
\hline
%\hline
Gas     & & 170 MeV/c & 440 MeV/c & MIPs\\
Mixture & & Pion     & Proton           \\ [0.5ex] 
\hline	% inserts single-line

% Entering 1st row 
 &clu/cm      &45.2$\pm$2.1 &96.6$\pm$3.5&40.0$\pm$2.0 \\[-1ex]
\raisebox{1.5ex}{Ar/C$_4$H$_{10}$} & \\[-1ex] 
\raisebox{1.5ex}{80/20}
\raisebox{1.5ex} &e$^-$/clu &2.13$\pm$0.12&2.12$\pm$0.11&2.11$\pm$0.11 %\\[-1ex]
\\[0.5ex] %adds vertical space 
\hline	% inserts single-line 

% Entering 2nd row 
  &clu/cm     &37.2$\pm$1.9&79.6$\pm$2.8&32.8$\pm$1.8  \\[-1ex] 
\raisebox{1.5ex}{Ar/C$_4$H$_{10}$} & \\[-1ex] 
\raisebox{1.5ex}{90/10}
\raisebox{1.5ex} &e$^-$/clu &2.14$\pm$0.12&2.12$\pm$0.11&2.12$\pm$0.10 %\\[-1ex]
\\[0.5ex] %adds vertical space 
\hline	% inserts single-line 

% Entering 3rd row 
%  &clu/cm      &3.7$\pm$0.7&7.9$\pm$0.9&3.3$\pm$0.6  \\[-1ex]
%\raisebox{1.5ex}{Helium} & \\[-1ex] 
%\raisebox{1.5ex}{100}
%\raisebox{1.5ex} &e$^-$/clu &2.38$\pm$0.14&2.31$\pm$0.13&2.36$\pm$0.13 %\\[-1ex]
%\\[0.5ex] %adds vertical space 
% [1ex] adds vertical space 
%\hline	% inserts single-line 

% Entering 3rd row 
  &clu/cm      &32.2$\pm$1.8&68.8$\pm$2.6&28.4$\pm$1.6  \\[-1ex]
\raisebox{1.5ex}{Ar/CO$_2$} & \\[-1ex] 
\raisebox{1.5ex}{70/30}
\raisebox{1.5ex} &e$^-$/clu &2.19$\pm$0.13&2.20$\pm$0.13&2.21$\pm$0.14 %\\[-1ex]
\\[0.5ex] %adds vertical space 
% [1ex] adds vertical space 
\hline	% inserts single-line 
\end{tabular} 
\caption{Primary ionization in the tested gas mixture for different 
         particle momentum. 
         The Ar/CO$_2$=70/30 gas mixture and MIPs are reported for 
         comparison.} % title name of the table
\label{tab:cluster} 
\end{table}

\subsection{Gas Gain Measurement}
\label{sec:gain}
\begin{figure}[!ht]
  \centering
    \includegraphics[width=7.5cm]{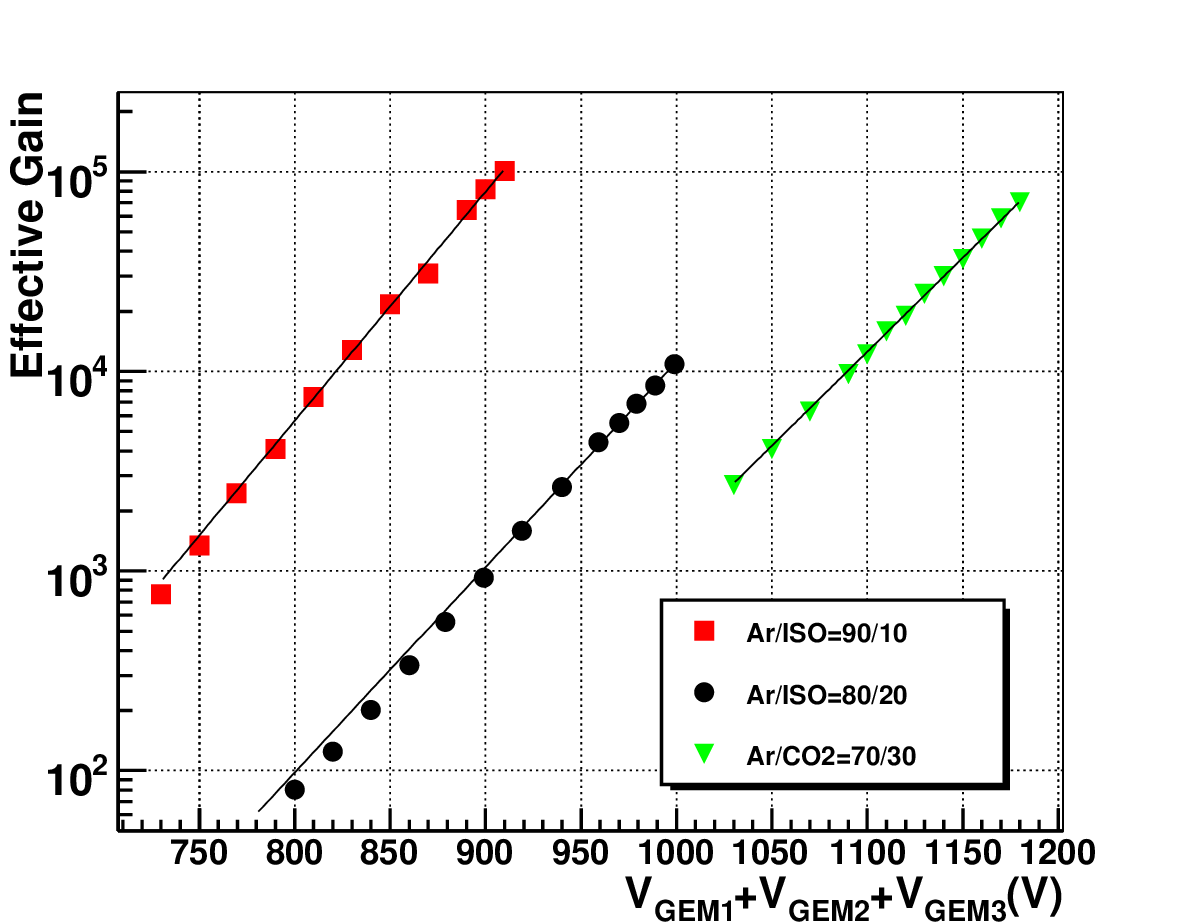}
    \caption{Effective gain as a function of
             the sum of the voltages applied on the three GEM
             foils for different gas mixture. The curve of the Ar/CO$_2$=70/30 
             gas mixture is reported for comparison.}
    \label{fig:gain}
\end{figure}

The effective gain of a triple-GEM
detector has been measured for different gas mixtures
using a high intensity 5.9 keV X-ray tube.
Effective gain value were obtained from the ratio of pad 
current with high voltage on the GEM foils, to current 
on the first GEM, with no high voltage on the GEM foils as 
shown in Fig.~\ref{fig:gain}.
For each gas mixture the detector has been
operated with field configurations optimising
the electron transparency~\cite{gain} and the ion feed-back.

\section{Detector Performances}
\label{sec:results}
%spiegare come facciamo l'analisi: determinazione T0
%(due pad all'estremo nel track fit e sposto
%la distribuzione dei residui per ogni pad)
%Ridefinisco il tempo delle pad da questo T0
%e procedo ricalcolando questa volta i residui
%togliendo dal fit la pad 
%

\subsection{Drift Velocity}
\label{sec:drift} 
\begin{figure}[h!]
  \centering
    \includegraphics[width=7.5cm]{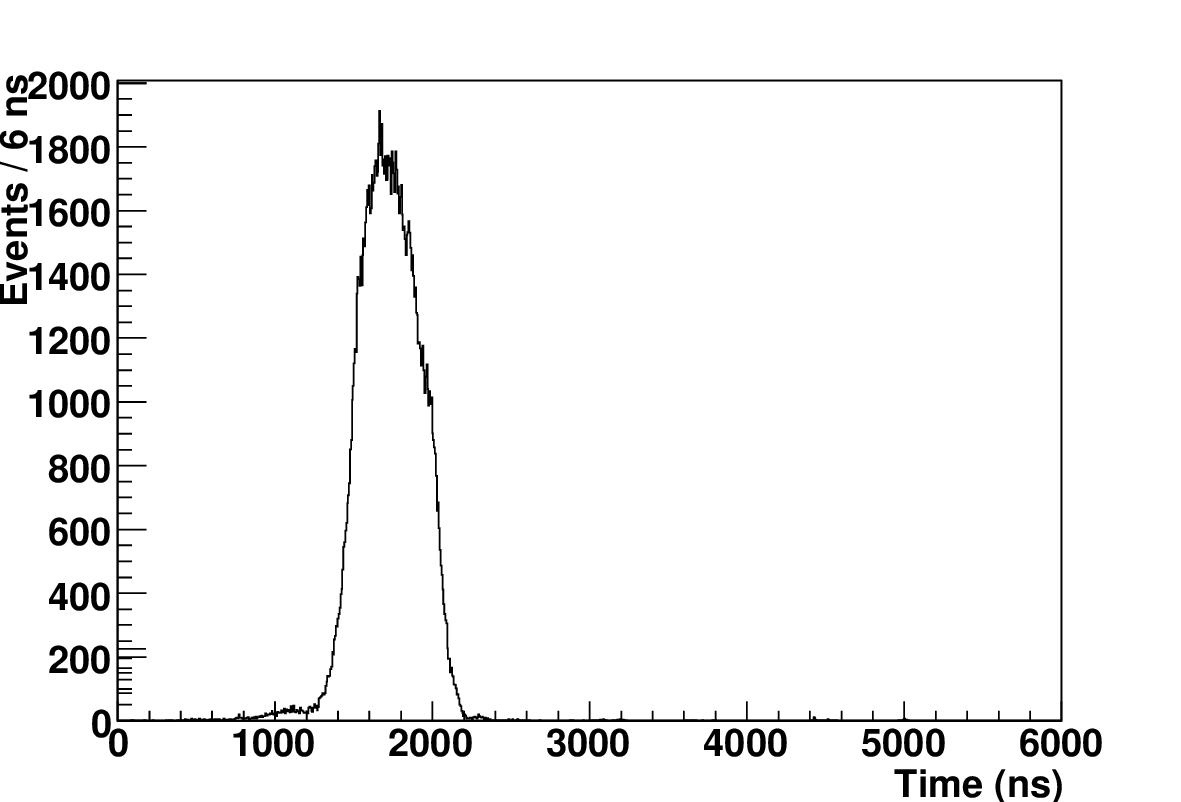}
    \caption{Drift-time distribution: Ar/C$_4$H$_{10}$= 90/10 gas mixture; 
             Drift Field= 200 V/cm; Gas Gain= 8$\times$10$^3$.}
    \label{fig:z0}
\end{figure}

The measurement of the drift velocity of the primary electrons 
in the drift gap for a fixed gas mixture is needed for 
the time-space relationship.\\
For this measurement the beam was shot perpendicular to the field cage.
Since it was not possible to measure with great accuracy the distance of 
the impact point of the beam on the detector field-cage with respect to 
the upper face of the first GEM,
% and since the vertical size of the beam 
%was larger than the S3 scintillator used in the trigger, 
the drift velocity 
has been calculated as the ratio between the S3 scintillator height 
(20$\pm$1 mm)
and the full width at half maximum of the drift-time distribution. The relative 
error on the drift velocity is less than 2 $\mu$m/ns.\\  
Fig.~\ref{fig:z0} shows an example of the drift-time distribution.
 It should be noted that the edge 
of that distribution is cut by the finite size of the S3 scintillator height.

%Fig.~\ref{fig:z0} shows an example of the drift-time distribution due to the
%vertical size of the beam spot. It should be noted that the edge 
%of that distribution is cut by the finite size of the S3 scintillator height (2 cm).\\
%The drift velocity has been calculated as the ratio between the scintillator height
%and the full width at half maximum of the Gaussian distribution. 
%

\begin{figure}[ht!]
  \centering
    \includegraphics[width=7.5cm]{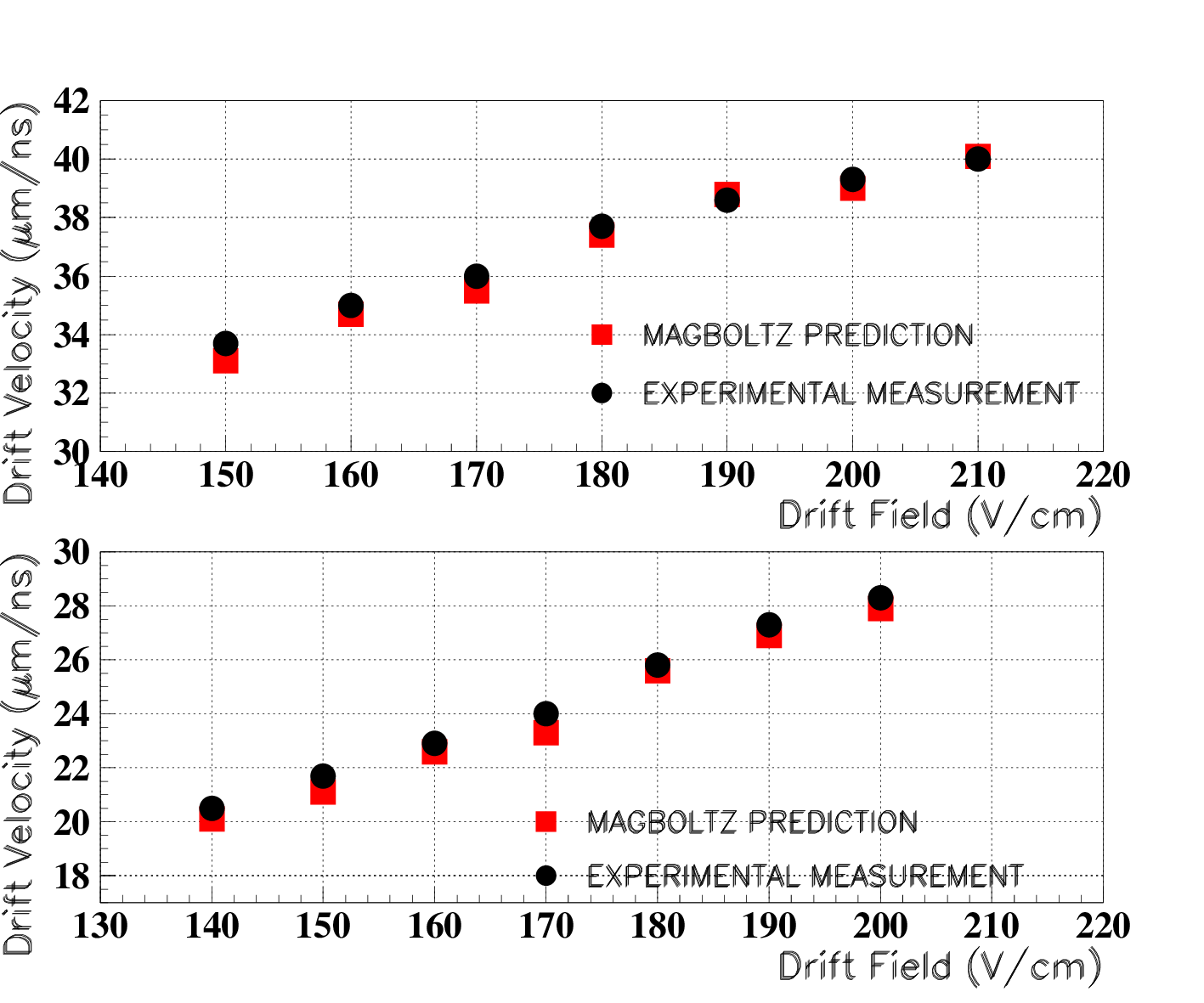}%drift1.eps
%    \includegraphics[height=3.cm, width=3.5cm]{drift1.eps}%
    %\qquad
%    \label{fig:drift1}
%    \qquad
%    \includegraphics[width=7.5cm]{drift2.eps}%
%    \includegraphics[height=3.cm, width=3.5cm]{drift2.eps}
     \caption{Measurements of the electron drift velocity as a function
      of the drift field. Magboltz predictions are also shown for comparison.
      Up: for the  Ar/C$_4$H$_{10}$= 90/10;
      Bottom:  Ar/C$_4$H$_{10}$= 80/20.}
      \label{fig:drift2}
\end{figure}

The drift velocity has been measured for different values of drift 
field and for 
the Ar/C$_4$H$_{10}$= 90/10 and Ar/C$_4$H$_{10}$=80/20 gas mixtures, 
respectively. 
The measurements are shown in Fig.~\ref{fig:drift2}, along with the 
corresponding Magboltz prediction~\cite{magboltz}. For comparison the 
drift velocity of the Ar/CO$_2$=70/30 is 3.5 $\mu$m/ns with a drift 
field of 150 V/cm.\\
In the following data analysis, the measured drift velocity has been 
used for the determination of the z-coordinate of 
the hits along the axis of the detector.

%
%\subsection{Longitudinal Diffusion}
%

\subsection{Efficiency}
\label{sec:efficiency}

\begin{figure}[ht!]
  \centering
    \includegraphics[width=6.9cm]{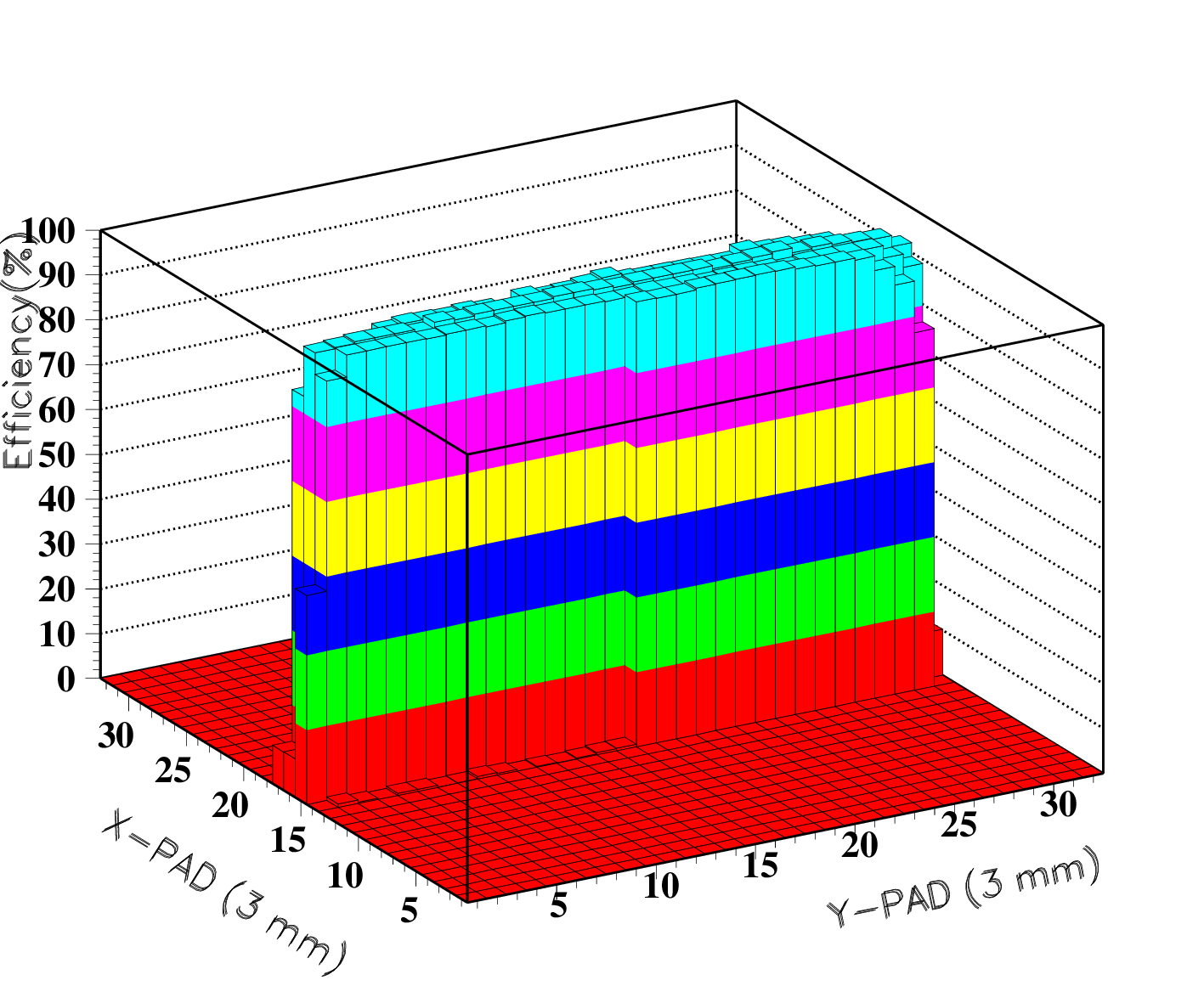}%
    \qquad
    \includegraphics[width=6.9cm]{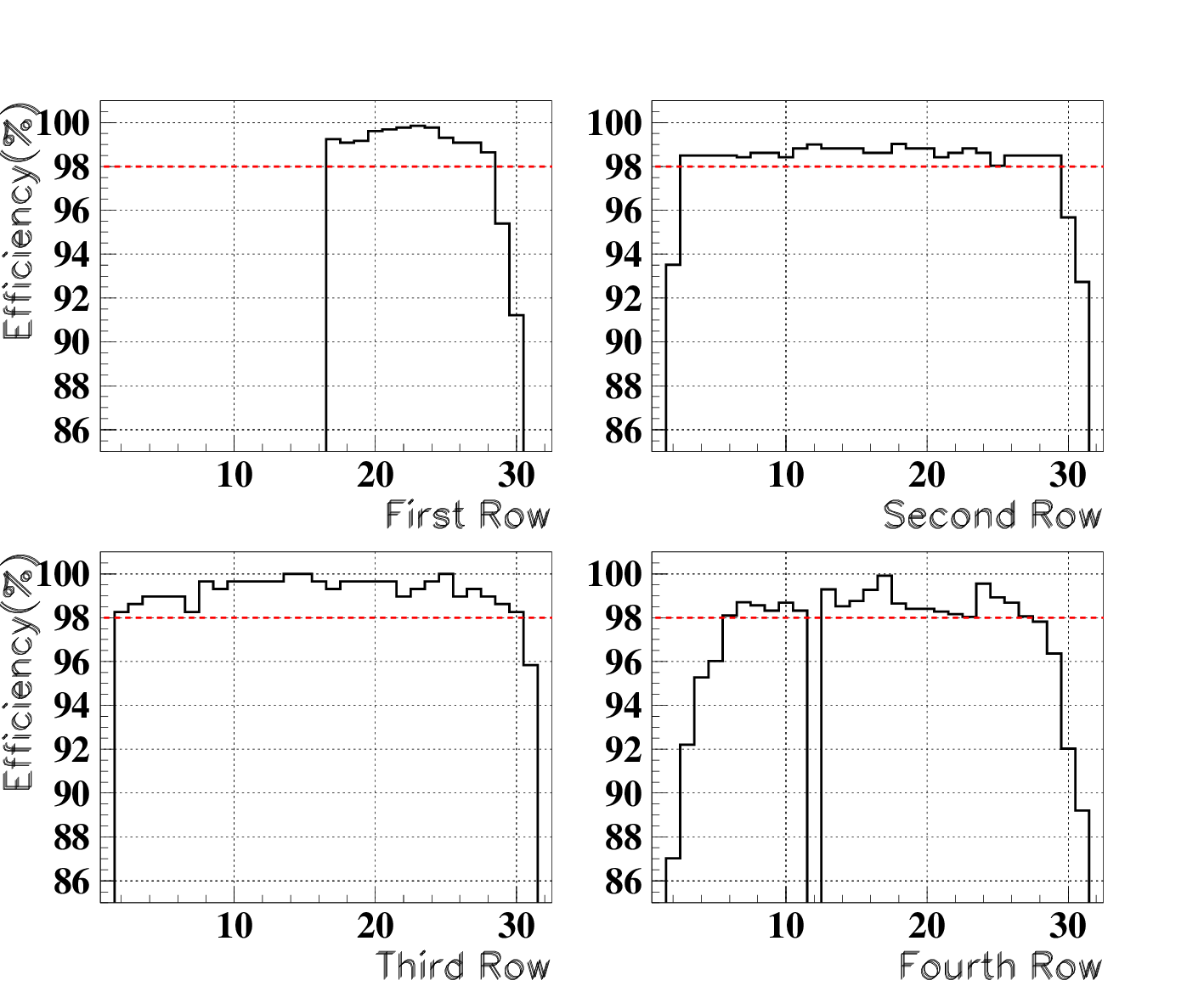}%
     \caption{Single pad row efficiency for the full detector (left) and 
              for each row (right). Each pad is $\sim$3$\times$3 mm$^2$.}
      \label{fig:efficiency}
\end{figure}

The single pad row detection efficiency has been evaluated considering 
the fraction of the hits in a single pad row with respect to a selected track.\\
%a single pad row with respect to a selected track; 
%The single pad row detection efficiency has been evaluated in two different
%methods: the first method is to considering the fraction of the hits in
%a single pad row with respect to a selected track; 
%the second one is to select track having hits in the 4$^{th}$ and the
%29$^{st}$ pad rows, and the fraction of the hits in the remaining pad row 
%was used as the detection efficiency.\\
%The two methods results to be equal if the single pad efficiency is greater 
%than $\sim$ 80\%, while differ at lower efficiency value. 
%Spiegare un pò meglio.\\
An example of the single pad efficiency for the full detector and for each 
row is shown Fig.~\ref{fig:efficiency}.

\begin{figure}[ht!]
  \centering
  \includegraphics[width=7.5cm]{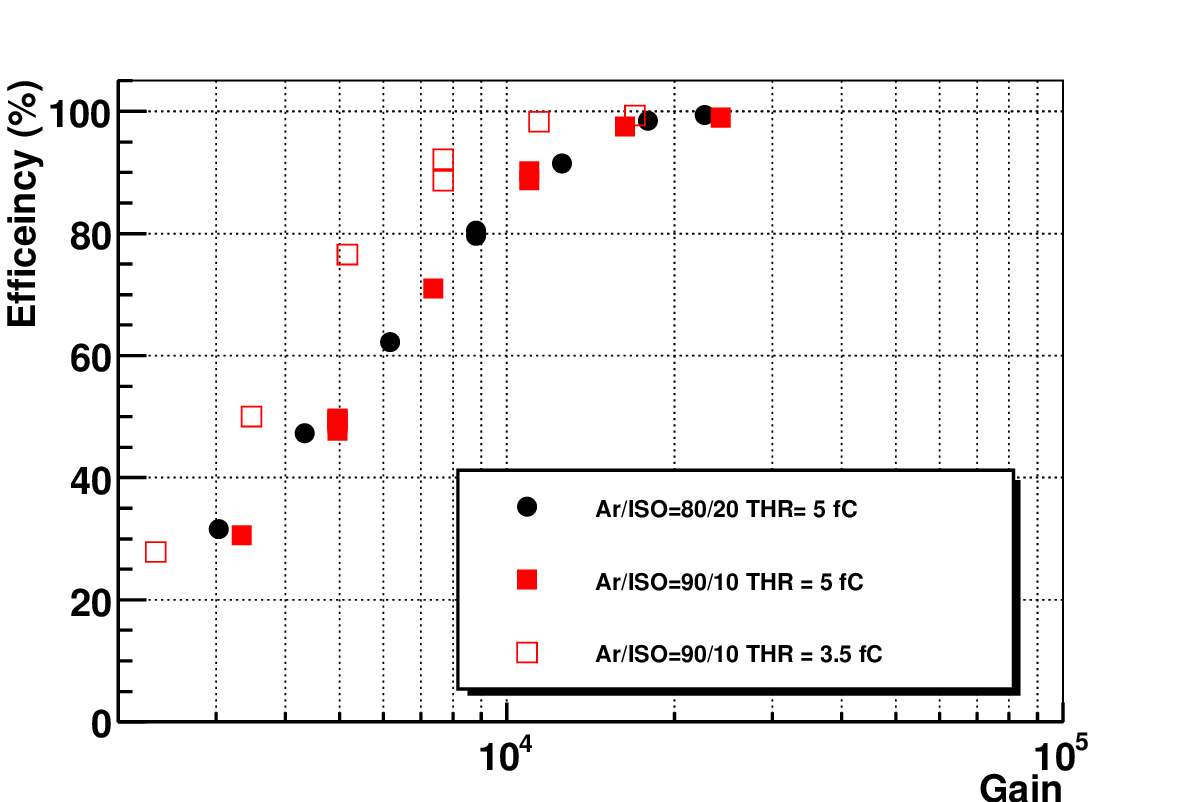}    
    \caption{Detection Efficiency for the Ar/C$_4$H$_{10}$= 80/20 and
     Ar/C$_4$H$_{10}$= 90/10 gas mixtures as a function of the gas gain 
     with a fixed drift field of 150 V/cm and 170 MeV/c pion beam.}
    \label{fig:effvsgain}
\end{figure}

It is worth noticing that most 
of the pads have a detection efficiency larger than 98\% except for 
the first 16 channels of the first row and the 11$^{st}$ channel of 
the fourth row that are dead. A low and/or not full detection efficiency 
in the first and last pads of each rows and a parabolic behaviour are clearly 
visible.\\ 
These effects will be discussed more in detail in Sec.~\ref{sec:field-cage}. 
Moreover, the first and the fourth rows suffer of a geometrical 
misalignment of the detector 
with respect to the trigger system which causes a small loss of hits.\\   
In the following, unless otherwise stated, the first and the last two pads
 of each row have been not considered in the selection of a track.\\
In Fig.~\ref{fig:effvsgain} the detection efficiency for the 
Ar/C$_4$H$_{10}$= 80/20 and Ar/C$_4$H$_{10}$= 90/10 gas mixtures 
as a function of the gas gain with a drift field of 150 V/cm and 170 MeV/c 
pions is shown. 
As expected the use of 3.5 fC front-end electronics threshold allows to
reach a full efficiency at lower values of gain 
with respect to the measurements performed with 5 fC of threshold.

\begin{figure}[ht!]
  \centering
  \includegraphics[width=7.5cm]{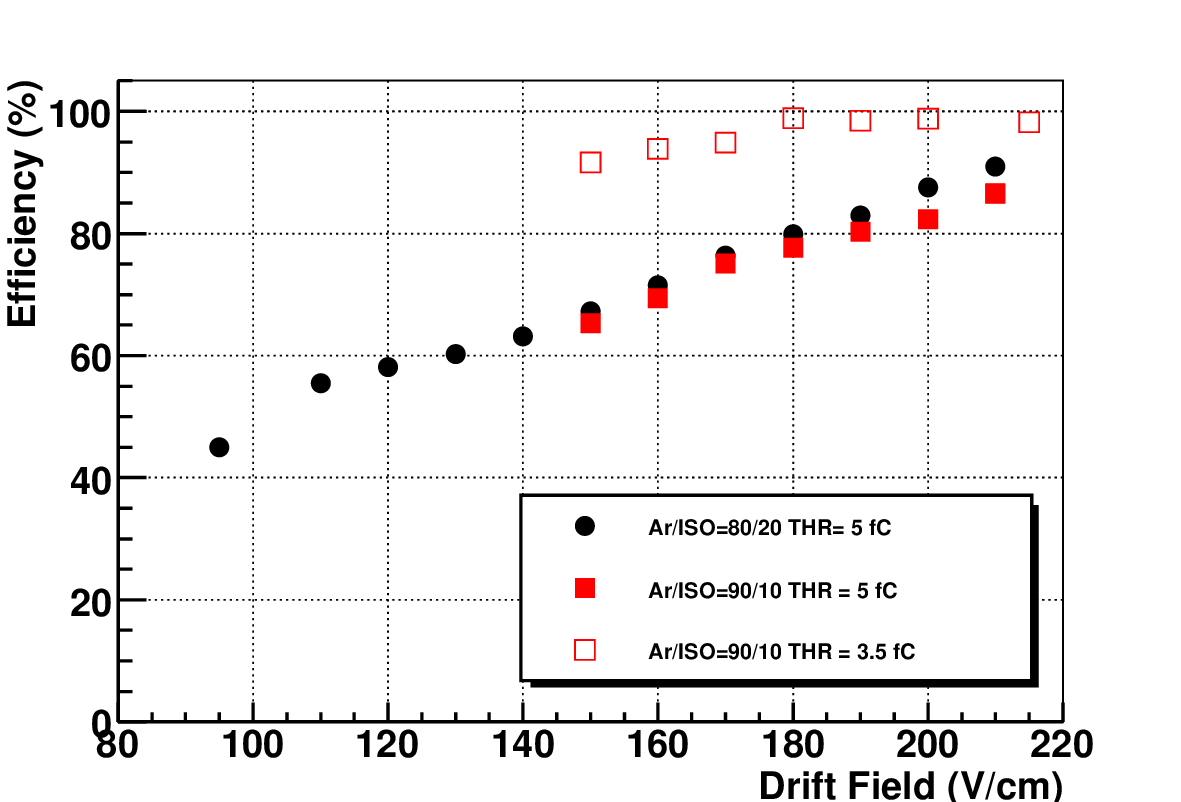}    
    \caption{Detection Efficiency for the Ar/C$_4$H$_{10}$= 80/20 and
     Ar/C$_4$H$_{10}$= 90/10 gas mixtures as a function of drift field with 
     a fixed gain of $\sim$8$\times$10$^3$ and 170 MeV/c pion beam.}
    \label{fig:effvsed}
\end{figure}

The increasing of drift field from 150 V/cm to 210 V/cm with a fixed 
gain of $\sim$ 8$\times$10$^3$  allows to increase the 
detection efficiency from $\sim$ 70\% to $\sim$ 90\% for both gas mixtures at 
5 fC threshold and from  $\sim$ 90\% to a full efficiency for 
Ar/C$_4$H$_{10}$= 90/10 
gas mixture at 3.5 fC threshold as shown in Fig.~\ref{fig:effvsed}.
This effect is due a greater 
collection efficiency of primary electrons in the first GEM when the drift 
field is increased.\\
Moreover, the different level of detection efficiency
between the two curves at 5 fC threshold is due to the higher number 
of primary electrons produced in the drift gap in the Ar/C$_4$H$_{10}$= 80/20 
gas mixture with respect to the Ar/C$_4$H$_{10}$= 90/10 one 
(see Tab.~\ref{tab:cluster}).

%\begin{figure}[ht]
%  \centering
%   \includegraphics[width=7.5cm]{effvsgain.eps}
%    \caption{Detection Efficiency for the  Ar/C$_4$H$_{10}$= 90/10 
%     gas mixture as a function of drift field with a gas gain of $\sim$10$^4$
%     and a front-end electronics threshold of about 5 fC.}
%    \label{fig:effvsdrift}
%\end{figure}
% 

\begin{figure}[h!]
  \centering
    \includegraphics[width=7.5cm]{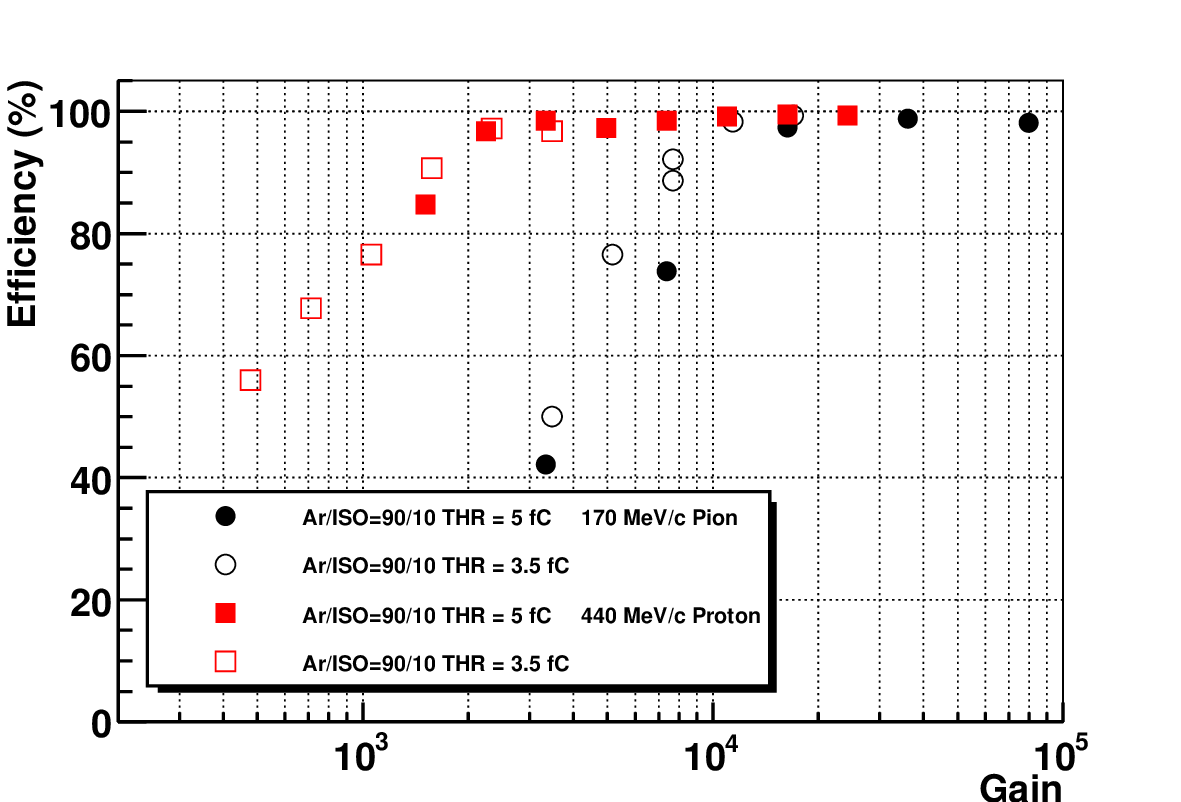}
    \caption{Detection Efficiency for protons of 440 MeV/c and pions
             of 170 MeV/c as a function of the gas gain. The gas mixture is 
             Ar/C$_4$H$_{10}$= 90/10 and the drift field is set to 150 V/cm.}
    \label{fig:effvspar}
\end{figure}

In Fig.~\ref{fig:effvspar} is shown the detection efficiency for 440 MeV/c
proton and 170 MeV/c pion beams crossing the detector. As expected, due to the 
higher ionization, protons allow to reach an efficiency plateau at lower
values of gas gain with respect to pions. The high value of primary ionization 
obtained with protons does not seem to affect the detection efficiency 
for the two threshold values, while for pions such effect does not occur, 
as explained before.    

\begin{figure}[ht!]
  \centering
    \includegraphics[width=7.5cm]{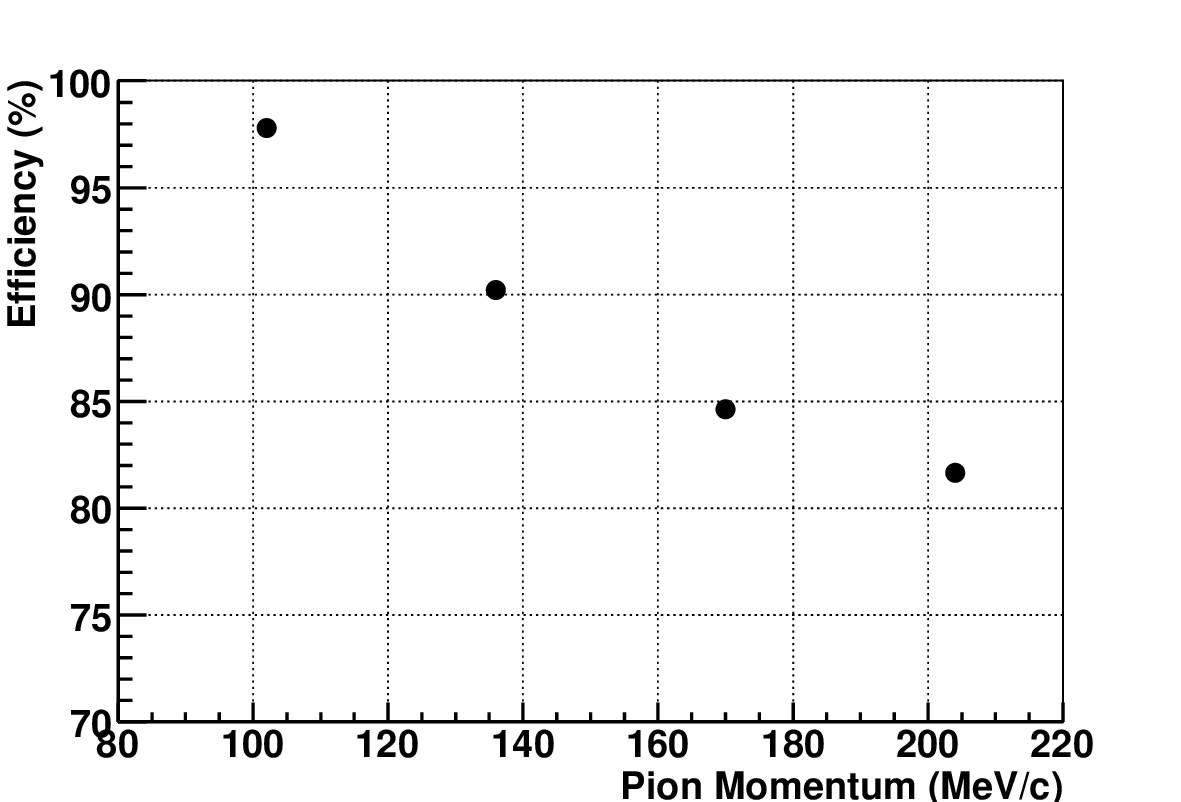}
    \caption{Detection Efficiency as a function of pion momentum 
             in Ar/C$_4$H$_{10}$= 80/20 gas mixture. The gain is set to 
             $\sim$ 8$\times$10$^3$ and the drift field is 190 V/cm.}
    \label{fig:effvsmom}
\end{figure}

The primary ionization effect has been also evaluated measuring the detector efficiency  
as a function of the pion beam momentum in the range between 
100 and 200 MeV/c (Fig.~\ref{fig:effvsmom}). As expected a higher 
efficiency value is obtained with a pion momentum of $\sim$ 100 MeV/c.

\subsection{Spatial Resolution}
\begin{figure}[ht!]
  \centering
    \includegraphics[width=7.5cm]{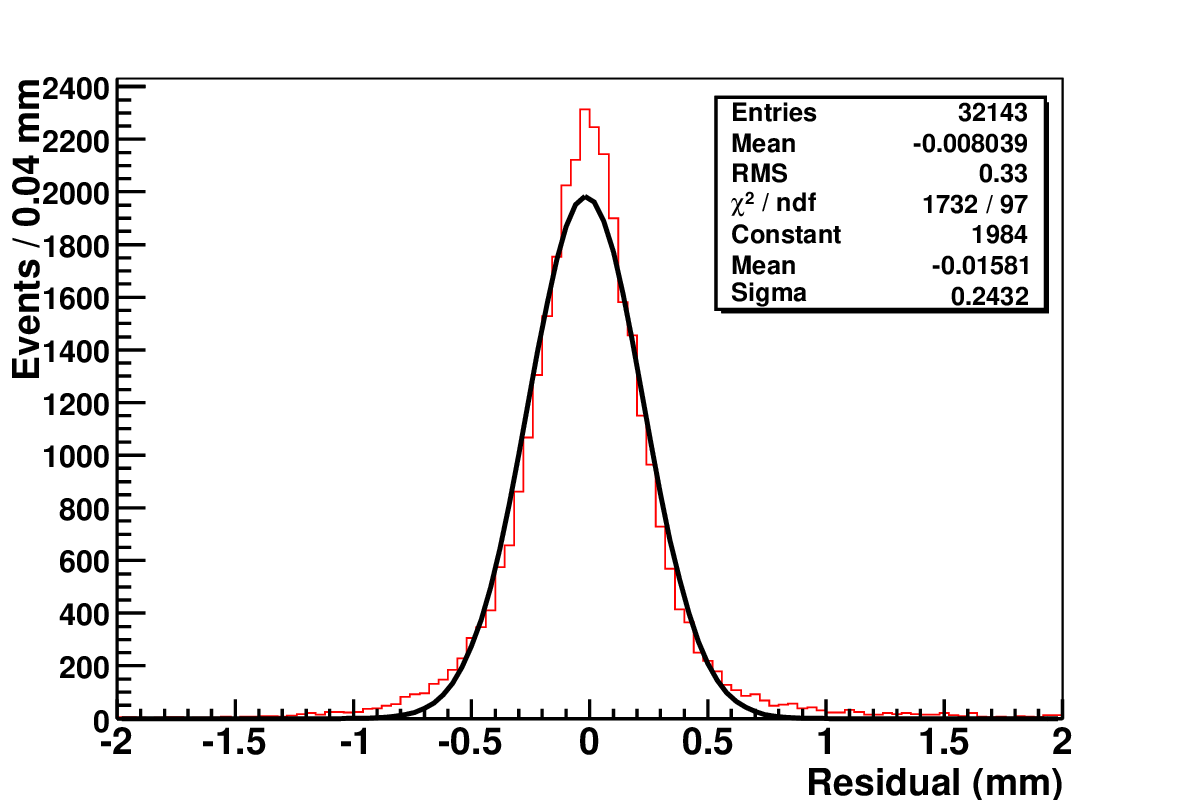}
    \caption{The best spatial resolution obtained with 440 MeV/c proton 
             and the Ar/C$_4$H$_{10}$= 90/10 gas mixture. The drift field is 
             set to 150 V/cm and the gas gain is $\sim$ 5$\times$10$^3$.}
    \label{fig:bestres}
\end{figure}

The spatial resolution in the drift direction was evaluated by the residuals 
between the hit position in a single pad row and a selected track.
The best spatial resolution obtained during the beam test is shown in 
Fig.~\ref{fig:bestres}.

\begin{figure}[ht!]
  \centering
    \includegraphics[width=7.5cm]{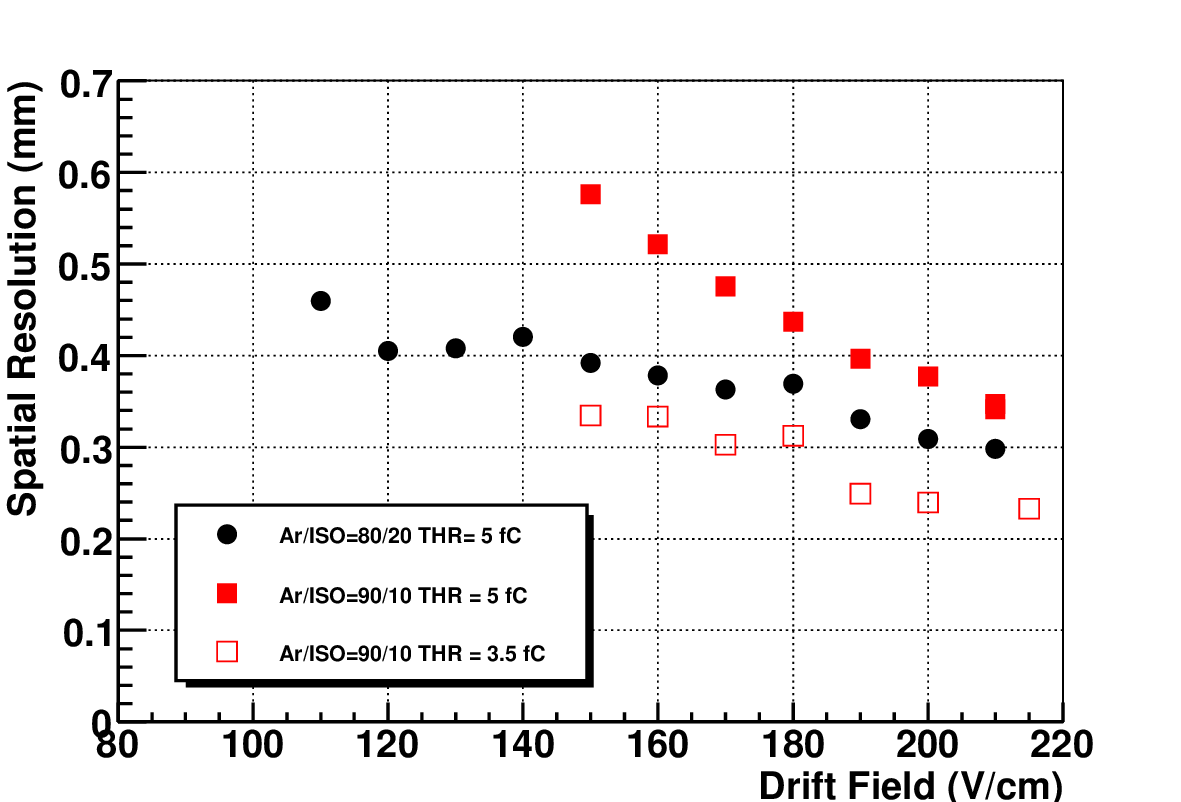}%res1.eps
     \caption{Spatial resolution in the drift direction for 
      the Ar/C$_4$H$_{10}$= 80/20 and Ar/C$_4$H$_{10}$= 90/10 
      gas mixtures as a function of
      the drift field with a fixed gain of $\sim$8$\times$10$^3$ and 170 MeV/c pions.}
      \label{fig:resvsed}
\end{figure}

Fig.~\ref{fig:resvsed} shows the spatial resolution in the drift direction 
for 170 MeV/c pions as a function of the drift field.
Since both the diffusion coefficient decreases and 
the collection efficiency of primary electrons in the first GEM increases
by increasing the drift field, a better spatial resolution is achieved 
for high drift field and low value of threshold.\\
Moreover, it should be noted that for 5 fC threshold value a better spatial 
resolution is reached with Ar/C$_4$H$_{10}$= 80/20 with respect to the 
Ar/C$_4$H$_{10}$= 90/10 gas mixture due to a lower diffusion coefficient.

\begin{figure}[ht!]
  \centering
    \includegraphics[width=7.5cm]{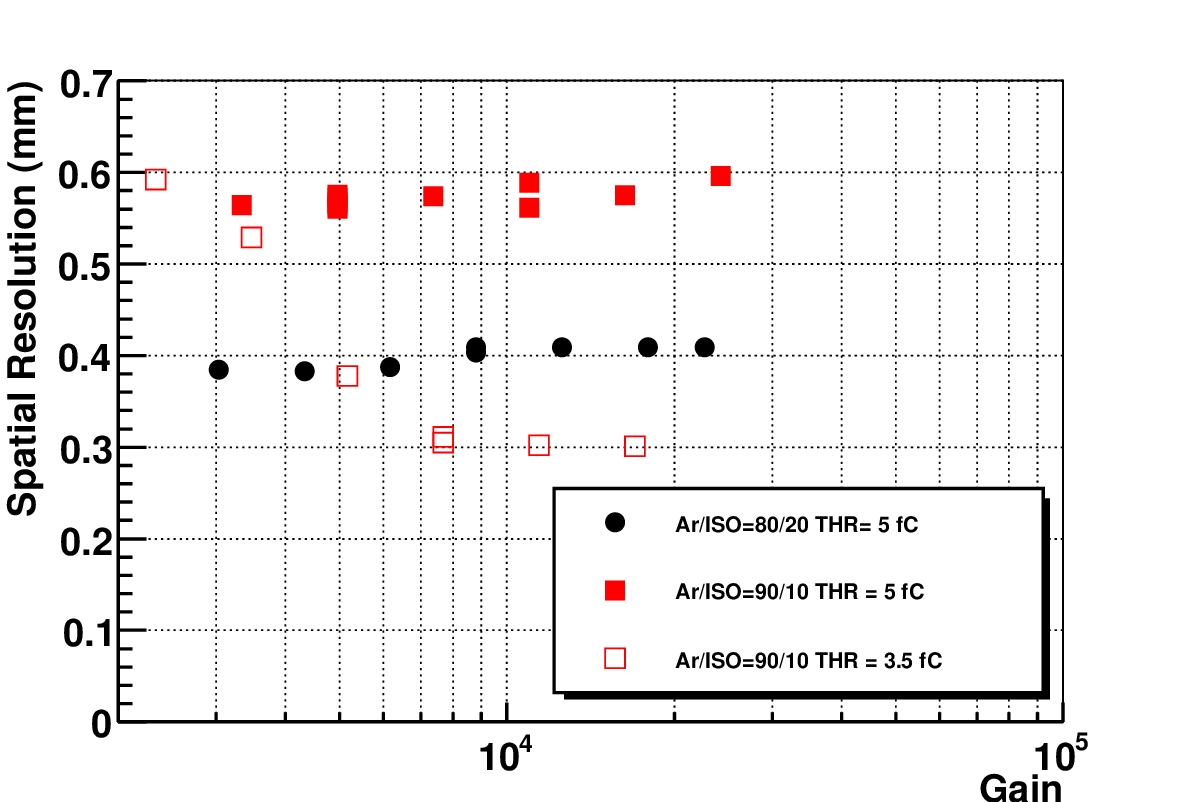}%res.eps
     \caption{Spatial resolution in the drift direction for 
      the Ar/C$_4$H$_{10}$= 80/20 and Ar/C$_4$H$_{10}$= 90/10 
      gas mixtures as a function of gas gain with a fixed drift field of 150 V/cm
      and 170 MeV/c pions.}
      \label{fig:resvsgain}
\end{figure}

The spatial resolution for 170 MeV/c pions as a function of the gas 
gain at fixed 
drift field is shown in Fig.~\ref{fig:resvsgain}.  
For the 3.5 fC threshold value, the spatial resolution 
decreases by increasing the detector gain due to a higher collection 
efficiency in the 
first GEM until it reaches a plateau region. On the contrary, for a high 
value of 
threshold and for both gas mixtures the spatial resolution seems to be not 
affected by the detector gain.
A possible explanation is that the collection efficiency in the first GEM 
increases as the detector gain increases until the signal is above the 
discrimination 
threshold. When this happens a better spatial resolution is achieved 
reaching a plateau 
when the signal is comparable with the electronic threshold.\\
Such explanation is confirmed by the fact that at very low gains 
($<$3$\times$10$^3$), 
the Ar/C$_4$H$_{10}$= 90/10 gas mixture reaches the same value of 
spatial resolution 
regardless of the used threshold.

\begin{figure}[h!]
  \centering
    \includegraphics[width=7.5cm]{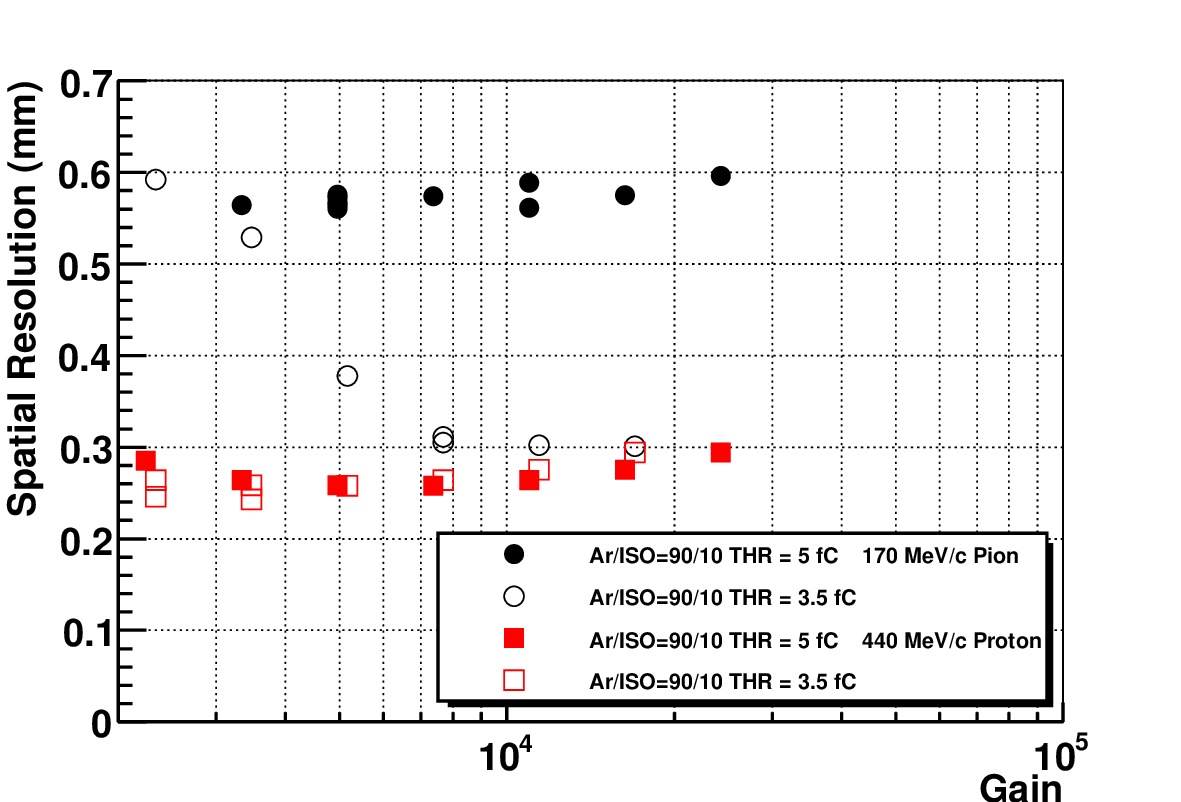}
    \caption{Spatial resolution for proton of 440 MeV/c and pion of 170 MeV/c 
             as a function of the gas gain. The gas mixture is 
             Ar/C$_4$H$_{10}$= 90/10 and the drift field is set to 150 V/cm.}
    \label{fig:resvspar}
\end{figure}

The effect of the ionizing particle on the spatial resolution has been 
evaluated comparing 440 MeV/c proton with 170 MeV/c pion as shown in 
Fig.~\ref{fig:resvspar}. As explained above, 
the spatial resolution seems not so sensible to a gain change with 
5 fC threshold 
value and regardless of ionization particle. Moreover, the larger 
ionization 
produced by protons with respect to pions allows to reach a better 
spatial resolution 
of about a factor 2.\\
At 3.5 fC of threshold, the spatial resolution for pions reaches 
at high gain the same 
level obtained with protons. This indicates that, together with the 
fact that there 
are no differences in the measurement of spatial resolutions with 
protons at different 
threshold valuesœôòëœôòë, the spatial resolution has reached a limit value for the 
Ar/C$_4$H$_{10}$= 90/10 gas mixture.

\begin{figure}[h!]
  \centering
    \includegraphics[width=7.5cm]{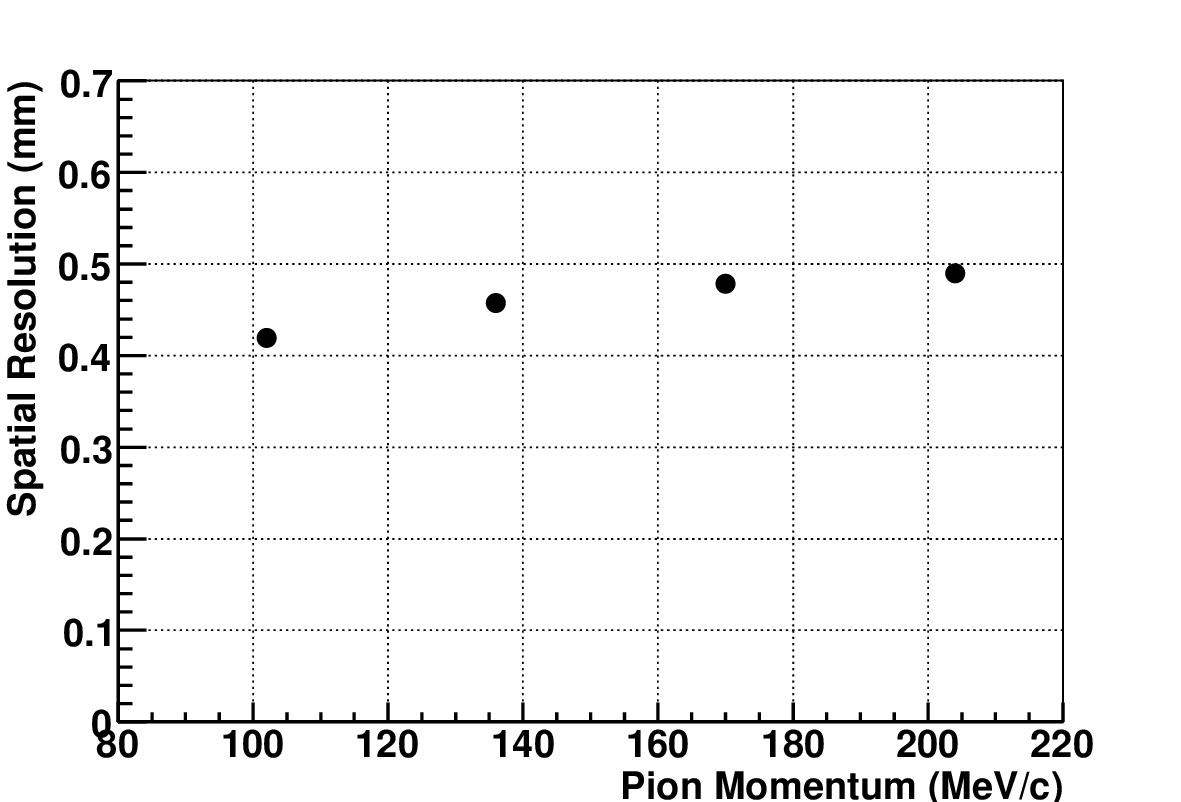}
    \caption{Spatial resolution as a function of pion momentum 
             in Ar/C$_4$H$_{10}$= 80/20 gas mixture. The gain is set to 
             $\sim$ 8$\times$10$^3$ and the drift field is 190 V/cm.}
    \label{fig:resvsmom}
\end{figure}

Fig.~\ref{fig:resvsmom} shows the spatial resolution as a function of pion beam momentum 
in the range between 100 MeV/c and 200 MeV/c. As expected, the spatial resolution 
results to be effected by low value of pion momentum.

\subsection{Time Over Threshold measurement}
\begin{figure}[ht!]
  \centering
    \includegraphics[width=7.5cm]{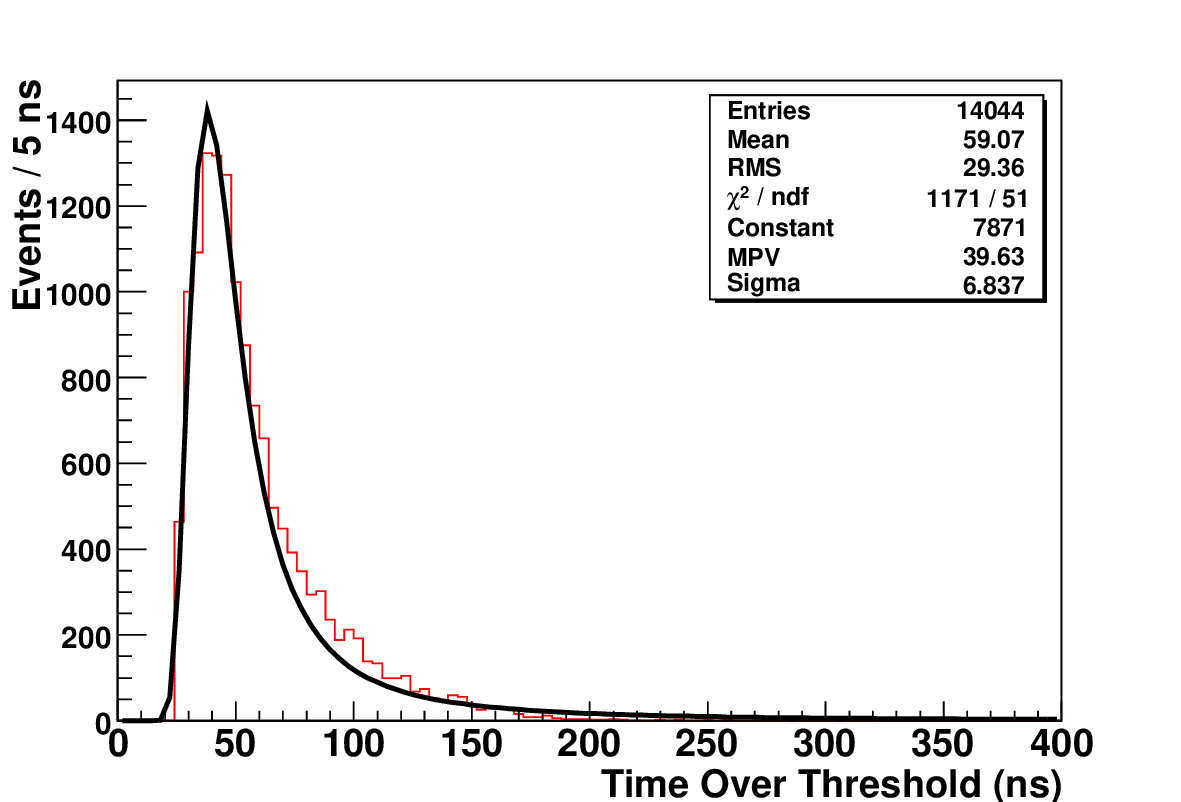}
    \caption{Distribution of the Time Over Threshold (TOT) measurements 
             along tracks with a 100 MeV/c pion beam. 
             The TOT distribution has been fitted with a Landau
             distribution.}
    \label{fig:charge}
\end{figure}

Timing functionality in the CARIOCA-GEM chip allows also the
measurement of charge, and therefore of ionization, for
individual channels. The measurement of signal pulse
width above a preset discriminator threshold may be used
to a good approximation as a determination of the charge
on each channel. From the collection of all channel measurements
the track ionization can be well determined.
As a confirmation of the ability of the CARIOCA-GEM chip to
measure simultaneously timing and ionization, the charge density 
along tracks has been determined using the Time Over Threshold (TOT)
measurement as shown in Fig.~\ref{fig:charge}.\\
%A precise determination of the primary track ionization
%has not been attempted, due to unknown efficiency factors
%(e.g., the charge not recorded when below the hit threshold).
It should be noted that the distribution of Fig.~\ref{fig:charge} follows 
a Landau distribution as expected. Moreover, the analysis has shown that this 
distribution is essentially independent of the track orientation.

\subsection{Particle Identification using dE/dx}
\begin{figure}[ht!]
  \centering
    \includegraphics[width=7.5cm]{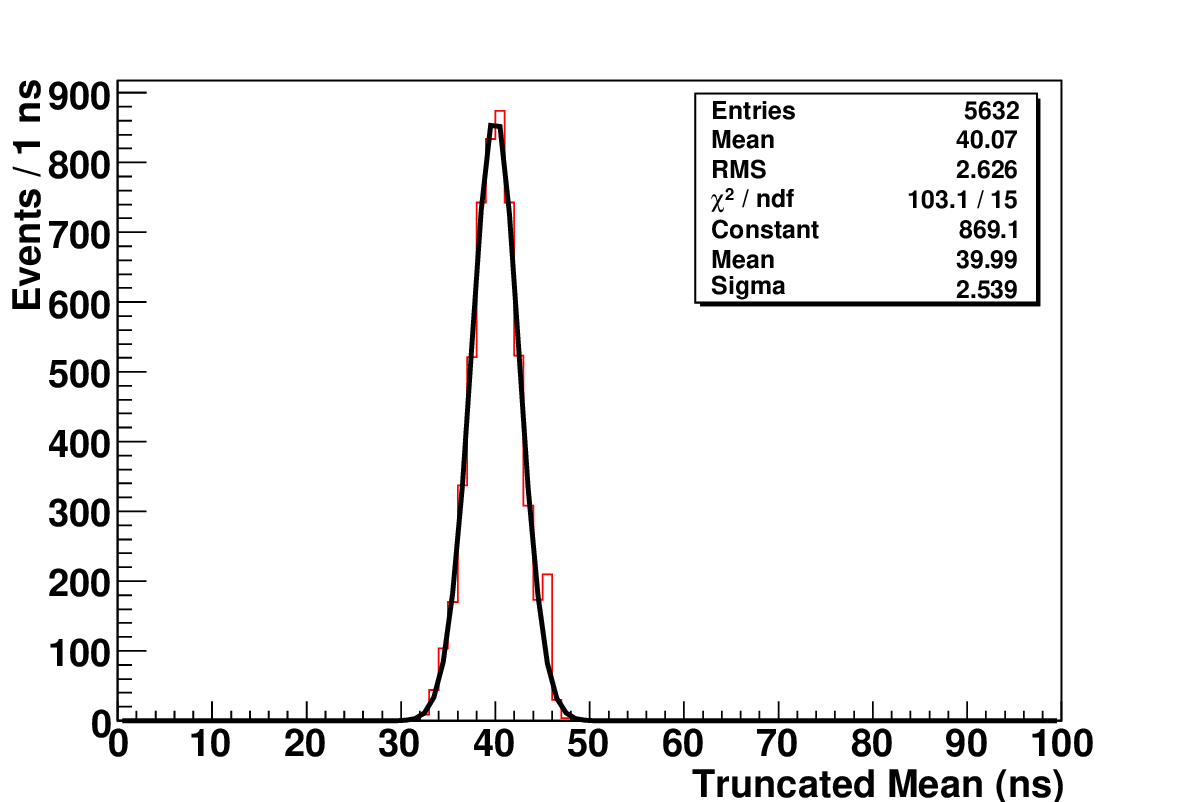}
    \caption{Truncated mean distribution in a sample of 30 hits
      obtained by removing the 60\% largest signals with a 100 MeV/c pion beam.}
    \label{fig:truncated}
\end{figure}

Energy loss, dE/dx, were measured for pions and protons to
evaluated the PID capability of our TPG prototype.\\
The dE/dx resolution was measured using the truncated technique. 
We observe that by accepting the 40\% lowest dE/dx values of the hits 
in the track we obtain the best resolution and we correctly reproduce the most 
probable value of the single pad charge distribution. For higher values of the
accepted fraction, the resolution gets worse due to inclusion of hits 
from the Landau distribution tail, while for smaller values the effect is
related to the loss in statistics.\\
Fig.~\ref{fig:truncated} shows the Gaussian behaviour of the 
resulting distribution when a cut of 60\% is applied.

\begin{figure}[ht!]
  \centering
    \includegraphics[width=7.5cm]{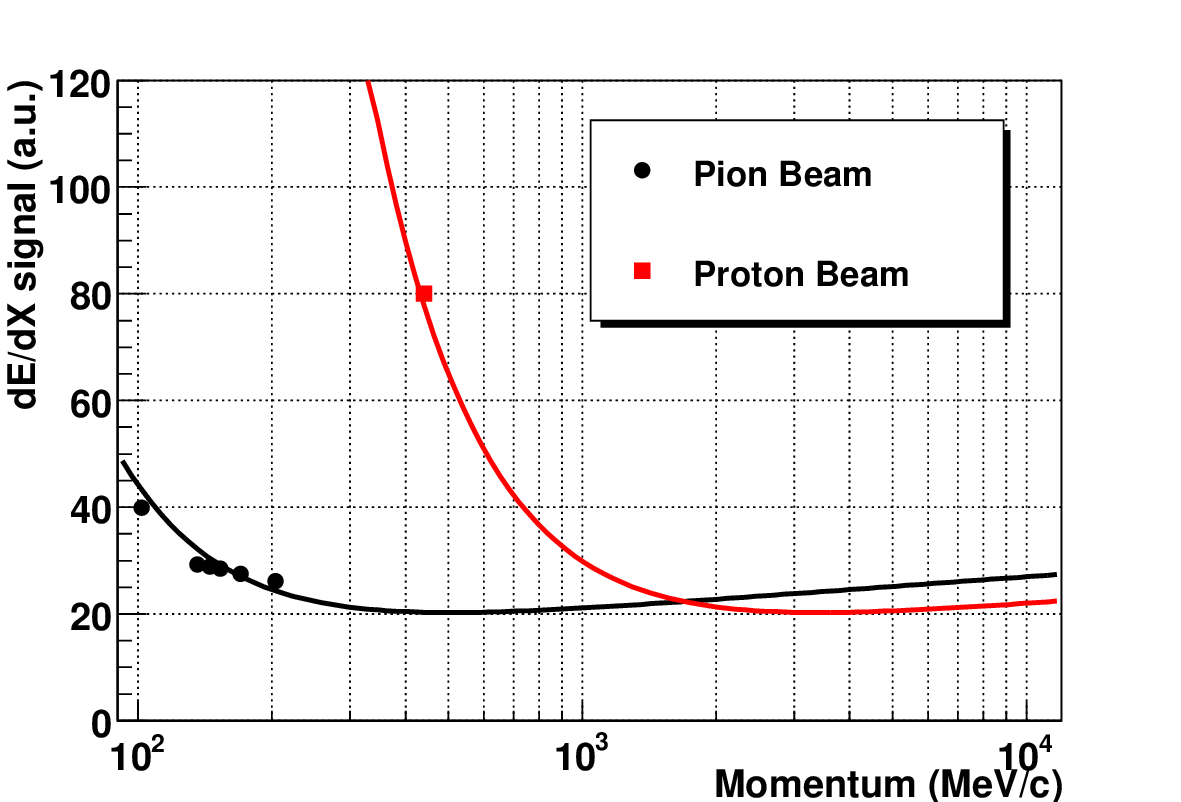}
    \caption{Measured dE/dx signal as a function of the momentum 
             for pion and proton beams with the Ar/C$_4$H$_{10}$= 80/20
             gas mixture.
             The gain is set to $\sim$ 8$\times$10$^3$ and the drift field 
             is 190 V/cm.}
    \label{fig:dEdX}
\end{figure}

By simultaneously measuring the momentum for proton and pion beams, 
by means of the time of flight (see Sec.~\ref{sec:setup}), and the deposited 
energy, by means of the mean value of the truncated distribution,
an estimation of the TPG prototype ability to identify the particle crossing 
the detector can be performed as shown in 
Fig.~\ref{fig:dEdX}.

\begin{figure}[h!]
  \centering
    \includegraphics[width=7.5cm]{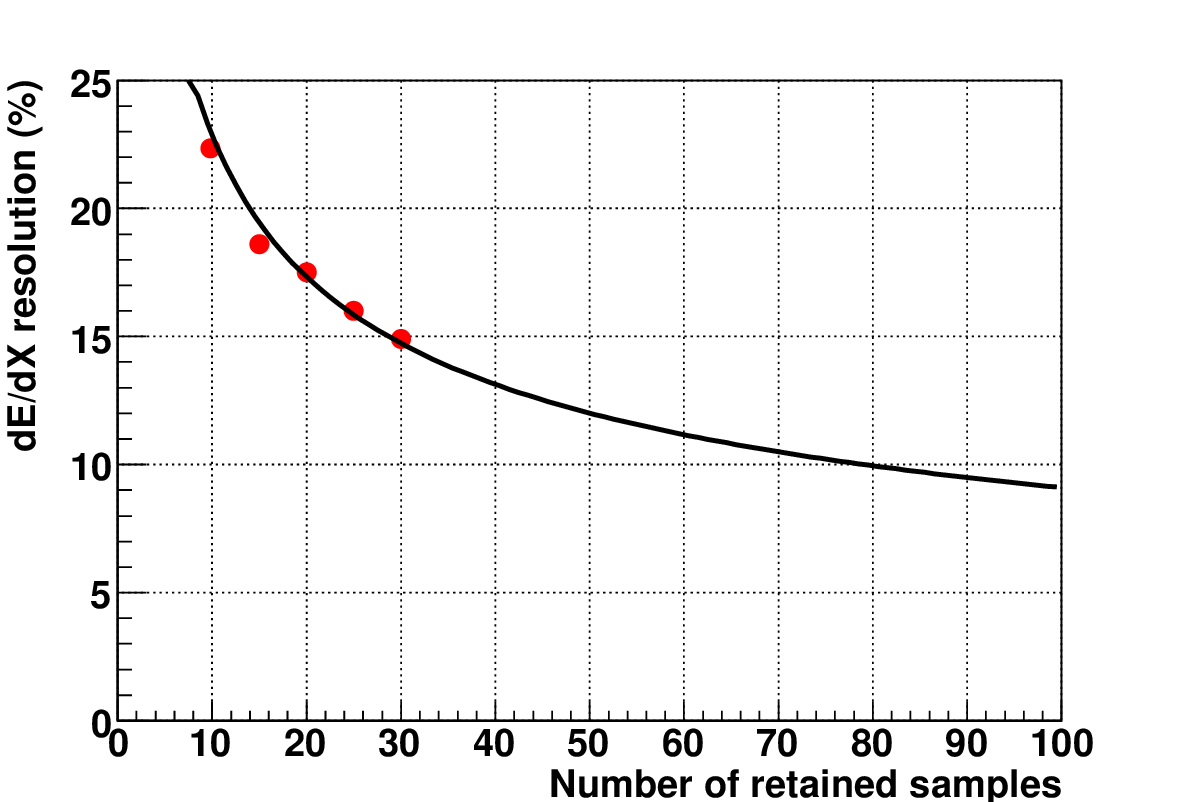}
    \caption{dE/dx resolution as function of the number of used hits
             with a truncated mean cut of 60\%. The curve is the expected dE/dx 
             resolution.}
    \label{fig:resE}
\end{figure}

For a fixed particle momentum, the dE/dx resolution function 
is usually parametrized~\cite{Cobb} as:
\begin{equation}
\sigma_{dE/dx} \propto N^a x^b
\end{equation}
where N the number of samples and x their length which in our case is
the pad dimension (Fig.~\ref{fig:resE}).
In the used Ar/C$_4$H$_{10}$= 80/20 gas mixture at atmospheric pressure, we 
measured a=-0.40$\pm$0.02 and b=-0.26$\pm$0.05.\\ 
%The single measurement resolution is (25.3$\pm$1.5)\%.\\
Assuming an average track length corresponding to 100 hits 
in the TPG-AMADEUS, with 100 MeV/c pion and 40\% of accepted fraction,
we expect to measure a dE/dx resolution of about 9\%. This energy resolution 
is comparable with that obtained with STAR TPC with a track length of more than 
67 cm~\cite{STAR}.\\ 
Moreover we expect to achieve, with such energy resolution and with 
a 100 MeV/c particle momentum, a K/$\pi$ and $\pi$/p separation power 
of 75 and 260, respectively.

% 
%In Fig.~\ref{fig:sep} particle separation are shown with an energy resolution 
%of 15\%, higher than to that achived with pions and protons beams. 
%As expected, hadron identification, and the relative separation, works well 
%in the low momentum range as foreseen in the AMADEUS experiment. On the contrary
%in the minimum ionization region, where hadrons carry momenta of few GeV/c and
%the ionization curves are very close, it is likely that the methode fails to
%discriminate the particles. 
%
%\begin{figure}[h!]
%  \centering
%    \includegraphics[width=8.0cm]{sep.eps}
%    \includegraphics[width=8.5cm]{c1.eps}
%    \caption{Separation power achievable with ionization measurements
%             assuming an energy resolution of 15\%.}
%    \label{fig:sep}
%\end{figure}
% 

\subsection{Edge Effect due to the Field Cage}
\label{sec:field-cage}
\begin{figure}[h!]
  \centering
    \includegraphics[width=7.8cm]{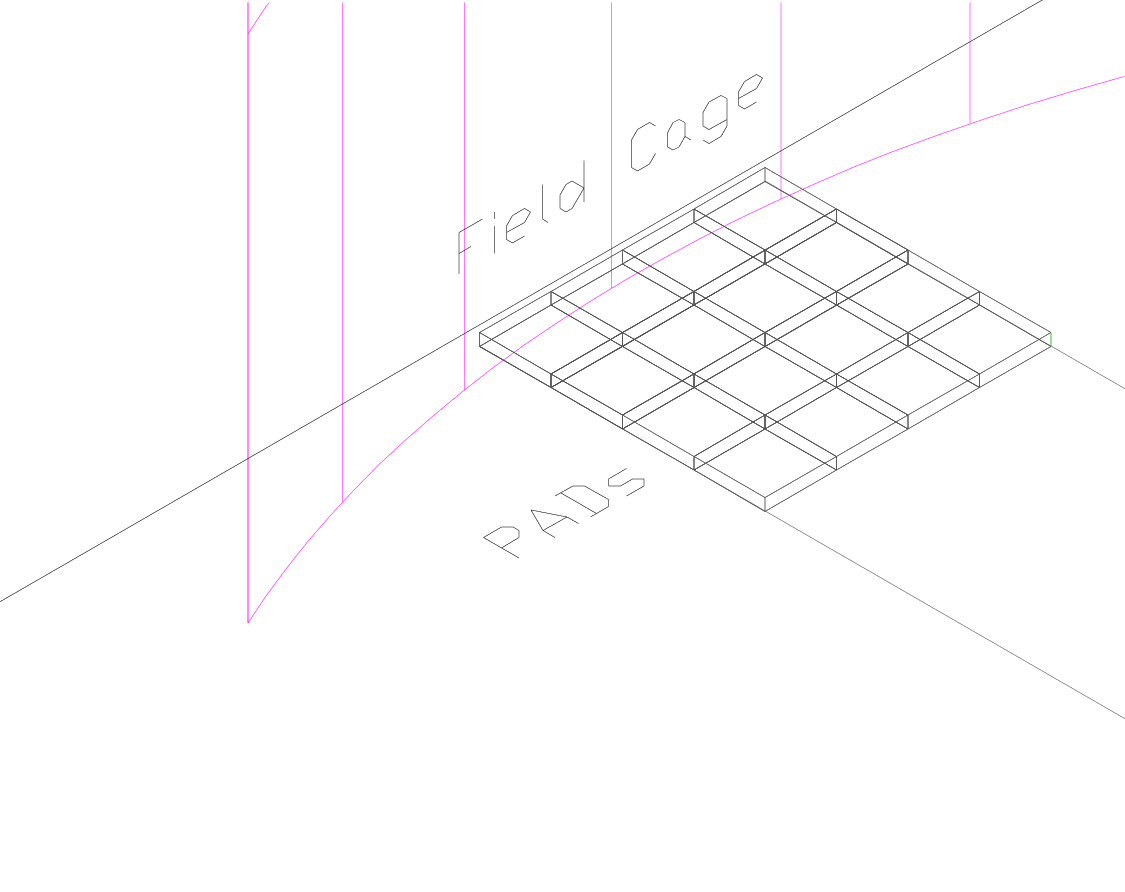}%field-cage.eps
    \caption{Drawing of the readout pads with the field cage wall.}
    \label{fig:field-cage-dr}
\end{figure}

\begin{figure}[h!]
  \centering
    \includegraphics[width=7.5cm]{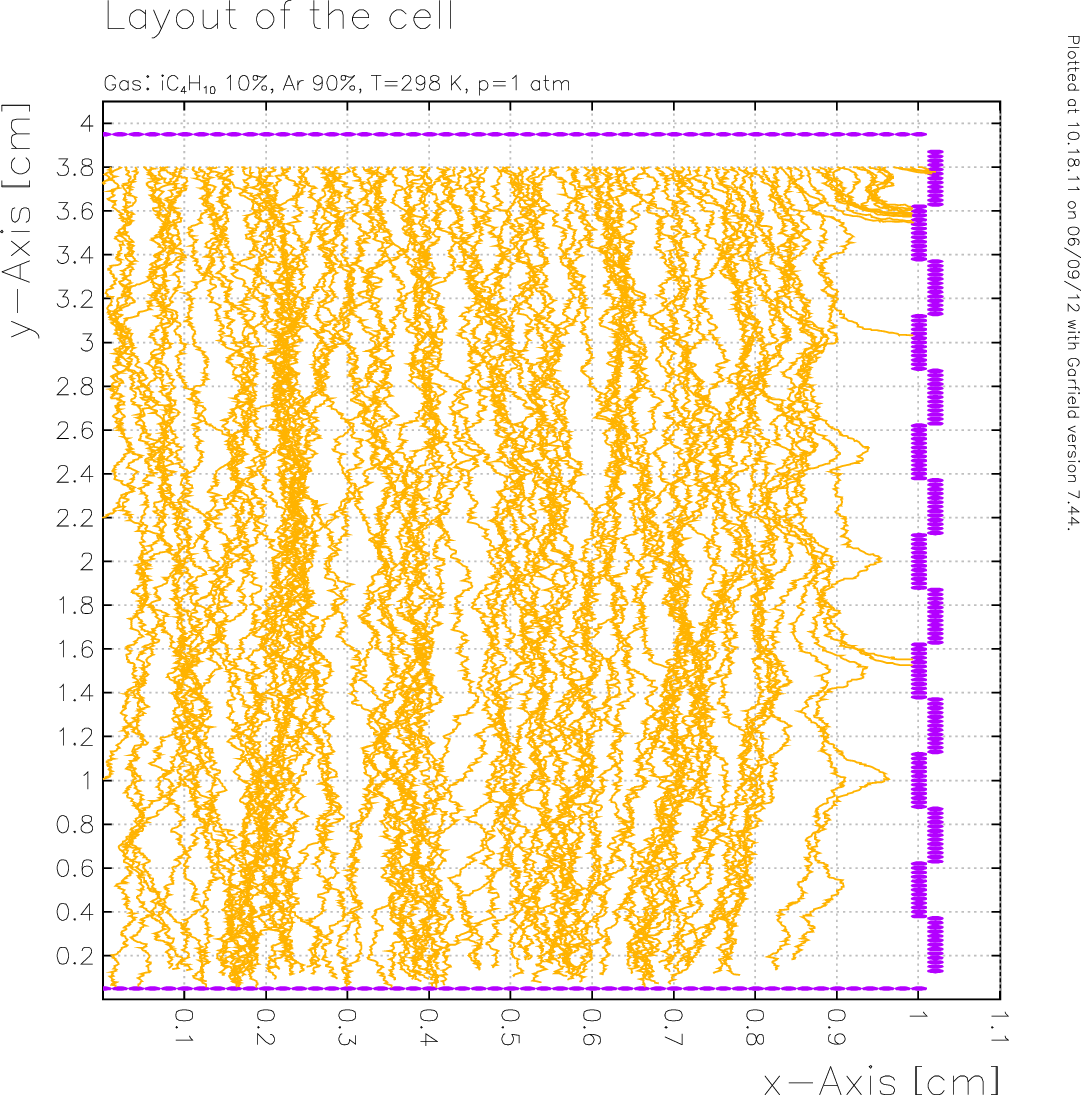}%field-cage.eps
    \caption{Garfield simulation of primary electrons in the drift gap.}
    \label{fig:field-cage}
\end{figure}

Low and/or not full detection efficiency has been measured on 
the edge of each pad rows as clearly visible in 
Fig.~\ref{fig:efficiency}.\\    
Such effects are due to different reasons:
\begin{itemize}
\item the length of each pad row is 102 mm while the effective diameter 
  of the cylindrical field-cage is 100 mm. 
  This means that the first and the last
  pad of each rows collect about 2/3 of the charge with 
  respect to the other pads of the row 
  (Fig.~\ref{fig:field-cage-dr});
\item electric distortion of the field in the drift gap mainly near the
      field-cage are not completely cured; 
\item the primary electrons produced in the drift gas and drifting toward the
      first GEM can be collected by the internal strips of the field-cage
       (Fig.~\ref{fig:field-cage}).  
\end{itemize}

In any case all these effects are fully reduced drifting away 
from the field-cage by $\sim$ 5 mm.


\hfill
\section{Conclusion}

The R\&D activity on TPG detector for the inner part of
the AMADEUS experiment has started at Laboratori Nazionali di Frascati (INFN). 
A TPG prototype, with a drift gap up to 15 cm, has been successfully 
produced and tested at the $\pi$M1 beam facility of the Paul 
Scherrer Institut with low momentum pion and proton beams.\\ 
The drift properties 
of various argon-isobutane gas mixtures have been measured and they result 
compatible with those simulated with Magboltz.\\
The measurement of the detector performances, in terms of detection 
efficiency and spatial resolution, as a function of the gas gain, drift field, 
front-end electronics threshold and particle momentum has been reported and 
discussed in detail.
A detection efficiency of 99\% and a resolution along the drift direction of 
240 $\mu$m have been achieved.\\ 
The dE/dx resolution has been measured for isobutane-based gas mixture 
applying a truncated mean of 60\%.
A good energy resolution, of about 25\% on a single measurement, 
will allow to reach an overall resolution of about 8\% for an average 
track length in AMADEUS-TPC.\\
Finally, the effect of the field-cage on the detector performances has been
also measured showing that it is necessary to move about 5 mm away from the 
field-cage.

\acknowledgments
%\section{Acknowledgements}
The authors would like to thank the coordinator of the PSI beam lines,
Dr. Konrad Deiters, for the excellent cooperation and support; 
%G. Corradi, D. Tagnani and C. Paglia for their support on the front-end 
%electronics, the Low voltage and High Voltage systems; 
M. Pistilli for his
suggestions during the prototype design phase and for his highly qualified 
technical assistance during its assembly.\\ 
Part of this work was supported by the European Community-Research 
Infrastructure Integrating Activity ``Study of Strongly Interacting Matter'' 
HadronPhysics 2 (HP2), Grant Agreement No. 227431, and
HadronPhysics 3 (HP3), Contract No. 283286, under
the Seventh Framework Programme of EU.

%Part of this work was supported by European Community Research Infrastructure 
%Interacting Activity ``Study of Strongly Interacting Matter'' (HadronPhysics2, 
%Grant agreement no. 227431) under the Seventh Framework Program of EU.\\
%The corresponding author and Dr. A. Romero Vidal are also thankful to Prof.
%C. Guaraldo for his support in this R\&D. 
%This study is supported by European Union Research in the Seventh Framework 
%Programme - Hadron Physics 3 (FP7-HP3).

\end{document}